\documentclass[authoryear,preprint,review,12pt]{elsarticle}

\usepackage[english]{babel}
\addto\captionsenglish{}

\usepackage{lineno}

\usepackage{amsmath}
\usepackage{amsfonts}
\usepackage{amssymb}

\usepackage{graphicx}
\graphicspath{{./}{./Figures/}}

\usepackage[table,dvipsnames]{xcolor}
\usepackage[normalem]{ulem}
\usepackage{bigstrut}
\usepackage{rotating}
\usepackage{multirow}
\usepackage{hhline}
\usepackage{units}

\usepackage{tikz}
\usepackage{pgfplots}
\usetikzlibrary{calc}
\usetikzlibrary{plotmarks}
\usetikzlibrary{patterns}
\usetikzlibrary{shapes}
\usetikzlibrary{external}
\usetikzlibrary{positioning}
\usepackage[english]{babel}

\usepackage{lipsum}               
\usepackage{pdflscape}

\definecolor{myGrey}{RGB}{240,240,240}

\def\XXint#1#2#3{{\setbox0=\hbox{$#1{#2#3}{\int}$}
     \vcenter{\hbox{$#2#3$}}\kern-.5\wd0}}

\usepackage{moresize}
\newcommand{\colnonsense}{a}
\newcommand{\substratDoveProbs}{\mathbf{s}}

\newcommand\clhdel{\bgroup\markoverwith{\textcolor{red}{\rule[.5ex]{2pt}{0.8pt}}}\ULon}

\usepackage{hyperref}
\hypersetup{
	pdftitle={Dominance, sharing, and assessment in an iterated Hawk--Dove game},
	pdfauthor={CL Hall, MA Porter, and MS Dawkins},
	colorlinks=true,              
}

\begin{document}
	
\begin{frontmatter}

\journal{Journal of Theoretical Biology}
\title{Dominance, Sharing, and Assessment in an Iterated Hawk--Dove Game}
\author[macsi]{Cameron L. Hall}
\author[ucla]{Mason A. Porter}
\author[oxfzoo]{Marian S. Dawkins}
\address[macsi]{Department of Engineering Mathematics, University of Bristol, UK \\
Department of Mathematics and Statistics, University of Limerick, Ireland \\ Mathematical Institute, University of Oxford, UK}
\address[ucla]{Department of Mathematics, University of California Los Angeles, USA \\ Mathematical Institute, University of Oxford, UK}
\address[oxfzoo]{Department of Zoology, University of Oxford, UK}


\begin{abstract}

	Animals use a wide variety of 
	strategies to reduce or avoid 
	aggression in conflicts over resources.
	These strategies range from sharing resources without outward signs of conflict to the development of dominance hierarchies, in which initial fighting is followed by the submission of subordinates.
	Although models
	have been developed to analyze specific strategies for resolving conflicts over resources,
	 little work has focused on trying to understand why particular strategies are more likely to arise in certain situations.
	In this paper, we use a model based on an iterated Hawk--Dove game to analyze how resource holding potentials (RHPs) and other factors affect whether sharing, dominance relationships, or other behaviours are evolutionarily stable.
	We find through extensive numerical simulations that sharing is stable only when the cost of fighting is low and the animals in a contest have similar RHPs, whereas dominance relationships are stable in most other situations.
	We also explore what happens when animals are unable to assess each other's RHPs without fighting, and we compare a range of strategies for this problem using simulations. We find (1) that the most successful strategies involve a limited period of assessment
	followed by a stable relationship in which fights are avoided and (2) that
	the duration of assessment 
	depends both on the costliness of fighting and on the difference between the animals' RHPs. 
	Along with our direct work on modeling and simulations, we develop extensive software to facilitate further testing; it is available at \url{https://bitbucket.org/CameronLHall/dominancesharingassessmentmatlab/}.	  
	
\end{abstract}




\end{frontmatter}



\section{Introduction}
\label{S:Intro}

When animals are in conflict over resources, the result can be anything from overt fighting \citep{ArcherBiologyAggression,HuntingfordAnimalConflict} to settling disputes by signaling \citep{CluttonBrock1979a}, formation of dominance hierarchies \citep{Bonabeau1999,Beacham1987,Drummond2006,Guhl1968,ODonnell1998}, or even resource sharing without any sign of overt conflict \citep{Wilkinson1984}.
To deal with this diversity of outcomes, a wide variety of game-theoretic models have been developed, including models that involve conditional strategies with assessment \citep{Parker1974,Parker1981,Enquist1990,Enquist1983,Payne1996}; iterated games, in which animals repeatedly encounter the same individuals \citep{Axelrod1981}; models based on simple learning rules \citep{Fawcett2010,grewal2013}; and models that include winner and loser effects (where winners are likely to keep winning and losers are likely to keep losing) \citep{Goessmann2000, Hsu2005, Kura2015, Kura2016, MestertonGibbons2016}.

This large variety of models often makes it difficult for biologists without detailed mathematical knowledge to understand the differences and similarities between these models. 
In general, models in which resources are ultimately divided unequally between `winners' and `losers' \citep{Eshel2001,Eshel2005,Fawcett2010,Hammerstein1981,Houston1991,Kura2015,Kura2016,MestertonGibbons2014,MestertonGibbons2016a} are often based on the Hawk--Dove framework that was described in \citet{MaynardSmith1979}, 
whereas models that concentrate on the evolution of sharing and other apparently paradoxical acts of cooperation \citep{Baek2017,Carvalho2016,Doebeli2005,Nowak1993a,Nowak1993,Nowak2012,Trivers2006} are often based on the Iterated Prisoner's Dilemma (IPD) framework that was described in \citet{AxelrodEvolCooperation}. This distinction seems to suggest that different models are needed for different outcomes.
Our aim in this paper is to show that, by contrast, a single model --- a modified version of the Hawk--Dove model that was developed by \citet{MaynardSmithEvolution} --- can predict the observed diversity of possible outcomes of animal conflict, including overt fighting, resource-sharing, dominance relationships, and other social structures.  
By making a small number of biologically-realistic modifications to the original Hawk--Dove model, we show that it is possible to derive many of the currently employed game-theoretic models of animal conflict and cooperation (including Conditional Hawk--Dove, Prisoner's Dilemma, and Snowdrift) and to explain the widespread occurrence of dominance hierarchies in animal societies. 

Our model, and the accompanying software, makes it easy to simulate many different games and to explore the effects of different assumptions and parameter values. With the present work, we hope to facilitate communication between mathematical modelers and field biologists, leading to a better understanding of why different animals resolve their conflicts in such different ways.


Our paper proceeds as follows. We present our model in Section \ref{S:Model}. We demonstrate how it shares several features with existing animal interaction models \citep{Eshel2001,Eshel2005,MestertonGibbons2014} and that, in certain limits, it reduces to a form of the IPD that was described in \citet{AxelrodEvolCooperation}. Our model addresses how differences in resource holding potential (RHP) can affect animals' optimal strategies while also providing a framework for modeling progressive assessment of RHPs by enabling animals' behaviours to change as they learn information about RHP through experience \citep{MaynardSmith1976,Parker1981,Enquist1983}. In Section \ref{S:Informed}, we investigate evolutionarily stable strategies (ESSs) for cases in which animals begin with complete knowledge of their RHP relative to those of their opponents. This forms the basis for Section \ref{S:Learning}, where we describe and analyze various strategies by which animals can use fights to learn about their RHP relative to those of their opponents. We conclude and discuss the implications of our results in Section \ref{S:Discussion}. Our software is available is at \url{https://bitbucket.org/CameronLHall/dominancesharingassessmentmatlab/}, and we encourage readers to pursue our ideas further and extend them.


\section{Model development}
\label{S:Model}


\subsection{An iterated Hawk--Dove game with winners and losers}
\label{S:HawkDove}

Consider two animals, $A$ and $B$, who interact with each other repeatedly in a contest for resources.
We construct a model that is based on the classical Hawk--Dove (HD) game (see Figure \ref{F:ClassicalHD}), which was developed to describe a single conflict between animals over a shareable resource \citep{MaynardSmith1973,MaynardSmith1979,MaynardSmithEvolution}, which we normalize to have a value of $1$. In the HD game, each animal has a choice between a Hawk strategy (in which it escalates the conflict) and a Dove strategy (in which it retreats from an escalated conflict). If two Doves encounter each other, they share a resource equally. If a Hawk and a Dove encounter each other, the Hawk takes the entire resource. If two Hawks encounter each other, a fight ensues. Assuming that each animal has an equal chance of winning, each animal receives a payoff of $(1-c)/2$, where $c$ represents the cost of fighting.

There are various systems for classifying $2 \times 2$ games according to the orders of the payoffs to the players \citep{Robinson2x2Games,Bruns2015}. For the classical HD game, the classification depends on the value of $c$. When $c > 1$, fights cost more than the value of the resource, so the 
classical HD game is an example of the `Snowdrift' game (also called the `Chicken' game\footnote{Although the term `Chicken' is used more widely than `Snowdrift' \citep{Robinson2x2Games,Bruns2015}, we use the latter to avoid any confusion between game-classification terminology and actual chickens.}). When $c \in (0,1)$, the HD game is an example of a `Prisoner's Dilemma' (PD). When $c = 1$, the classical HD game is not a strict ordinal game, as there are multiple ways for a player to obtain a payoff of $0$. Such `games with ties' \citep{Bruns2015} are more complicated to analyze than strict ordinal games. They require model parameters to take specific values, so they are `non-generic' \citep{BroomGameTheoreticalBiology} and are thus unlikely to occur in practice. Throughout this paper, we concentrate on ordinal games.

\begin{figure}
	\begin{center}
			\tikzsetnextfilename{HawkDoveMaynardSmith}
	\begin{tikzpicture}
\coordinate (halfVertGap) at (0,0.7);
\coordinate (halfHorGap) at (1.4,0);
\node[align=center] (HDPayoff) at (0,0) {$(1,0)$};
\node[align=center] (HHPayoff) at ($(halfHorGap)+(halfHorGap)$) {$\left(\tfrac{1-c}{2}, \tfrac{1-c}{2}\right)$};
\node[align=center] (DDPayoff) at ($(halfVertGap)+(halfVertGap)$) {$\left(\tfrac{1}{2},\tfrac{1}{2}\right)$};
\node[align=center] (DHPayoff) at ($(HHPayoff) + (DDPayoff)$) {$\left(0,1\right)$};
\coordinate (topLeft) at  ($(DDPayoff) - (halfHorGap) + (halfVertGap)$);
\coordinate (topRight) at ($(DHPayoff) + (halfHorGap) + (halfVertGap)$);
\coordinate (btmLeft) at  ($(HDPayoff) - (halfHorGap) - (halfVertGap)$);
\coordinate (btmRight) at ($(HHPayoff) + (halfHorGap) - (halfVertGap)$);
\coordinate (topMid) at   ($(DDPayoff) + (halfHorGap) + (halfVertGap)$);
\coordinate (btmMid) at   ($(HDPayoff) + (halfHorGap) - (halfVertGap)$);
\coordinate (midLeft) at  ($(HDPayoff) - (halfHorGap) + (halfVertGap)$);
\coordinate (midRight) at ($(HHPayoff) + (halfHorGap) + (halfVertGap)$);
\draw (topLeft) rectangle (btmRight);
\draw (topMid) -- (btmMid);	
\draw (midLeft) -- (midRight);

\coordinate (DoveA) at ($(DDPayoff) - (halfHorGap)$);
\coordinate (DoveB) at ($(DDPayoff) + (halfVertGap)$);
\coordinate (HawkA) at ($(HDPayoff) - (halfHorGap)$);
\coordinate (HawkB) at ($(DHPayoff) + (halfVertGap)$);
\node[above=6pt of DoveB,anchor=base] {Dove};
\node[above=6pt of HawkB,anchor=base] {Hawk};
\node[left=0.8cm of DoveA,anchor=center] {Dove};
\node[left=0.8cm of HawkA,anchor=center] {Hawk};
\node[left=2cm of midLeft,rotate=90,align=center,anchor=center] (AnimalA) {Animal $A$};
\node[above=18pt of topMid] (AnimalB) {Animal $B$};
\end{tikzpicture}
	\end{center}
	\caption{Payoffs to animals $A$ and $B$ in a classical Hawk--Dove game where the value of the resource is $1$ and the cost of fighting is $c$. 
}
	\label{F:ClassicalHD}
\end{figure}

The classical HD game and the IPD game are both inadequate for describing social interactions between animals. 
One key issue is that both of these games assume that there are no differences in the payoff matrices for the two animals. Without modification, neither game takes into account that one animal may have a larger RHP than the other, or that one animal may place a higher value on a disputed resource. 
Another issue is that neither game includes any element of assessment, where an animal uses information about its opponent from signals, resource ownership, or past experience to guide its behaviour. 
This is an important issue, as it is well-established that conditional strategies with assessment are far better models of animal conflict than strategies that do not involve assessment \citep{Parker1974}. 

In our model, we make three important modifications to the classical HD game:
\begin{enumerate}
\item{We study an iterated game in which each stage is an HD game.}
\item{We assume that there is always an explicit winner and loser in any Hawk--Hawk interaction, rather than supposing that each animal obtains an identical payoff.} 
\item{We assume that Hawk--Hawk conflicts have a biased outcome, in that one animal is more likely than its opponent to win the fight.}
\end{enumerate}
In Figure \ref{F:ModifiedHD}, we show our modified HD game in normal form. 

Our first modification is to consider an iterated game of HD stage games to model repeated interactions of two animals from the same social group. We assume that each HD interaction is identical, such that animals do not get stronger or weaker over time, and that the available resource has the same value in each stage game.

In our second modification, we assume that each Hawk--Hawk interaction has a winner and a loser, instead of each animal receiving an identical payoff. Specifically, we assume that there is a cost $c_W > 0$ of winning a fight and a cost $c_L > 0$ of losing a fight, such that the winner obtains a payoff of $1 - c_W$ and the loser obtains a payoff of $-c_L$. Additionally, $1 - c_W > -c_L$, as the payoff from winning a fight is greater than the payoff from losing it.

In our third modification, we suppose that one animal is more likely to win a Hawk--Hawk fight than its opponent. Specifically, we assume a fixed probability $p_A$ that animal $A$ beats animal $B$. Consequently, the expected payoff for animal $A$ from a Hawk--Hawk fight is $p_A (1 - c_W + c_L) - c_L$, and the expected payoff for animal $B$ is $1 - c_W - p_A(1 - c_W + c_L)$.

\begin{figure}
\begin{center}
	\tikzsetnextfilename{HawkDoveModified}
	\begin{tikzpicture}
	
	\coordinate (halfVertGap) at (0,0.8);
	\coordinate (halfHorGap) at (1.8,0);
	\node[align=center] (HDPayoff) at (0,0) {$(1,0)$};
	\node[align=center] (HHPayoff) at ($(halfHorGap)+(halfHorGap)$) {$(1-c_W,-c_L)$ \\ or $(-c_L,1-c_W)$};
	\node[align=center] (DDPayoff) at ($(halfVertGap)+(halfVertGap)$) {$(\tfrac{1}{2},\tfrac{1}{2})$};
	\node[align=center] (DHPayoff) at ($(HHPayoff) + (DDPayoff)$) {$(\tfrac{1}{2},\tfrac{1}{2})$};
	\coordinate (topLeft) at  ($(DDPayoff) - (halfHorGap) + (halfVertGap)$);
	\coordinate (topRight) at ($(DHPayoff) + (halfHorGap) + (halfVertGap)$);
	\coordinate (btmLeft) at  ($(HDPayoff) - (halfHorGap) - (halfVertGap)$);
	\coordinate (btmRight) at ($(HHPayoff) + (halfHorGap) - (halfVertGap)$);
	\coordinate (topMid) at   ($(DDPayoff) + (halfHorGap) + (halfVertGap)$);
	\coordinate (btmMid) at   ($(HDPayoff) + (halfHorGap) - (halfVertGap)$);
	\coordinate (midLeft) at  ($(HDPayoff) - (halfHorGap) + (halfVertGap)$);
	\coordinate (midRight) at ($(HHPayoff) + (halfHorGap) + (halfVertGap)$);
	\draw (topLeft) rectangle (btmRight);
	\draw (topMid) -- (btmMid);	
	\draw (midLeft) -- (midRight);
	
	\coordinate (DoveA) at ($(DDPayoff) - (halfHorGap)$);
	\coordinate (DoveB) at ($(DDPayoff) + (halfVertGap)$);
	\coordinate (HawkA) at ($(HDPayoff) - (halfHorGap)$);
	\coordinate (HawkB) at ($(DHPayoff) + (halfVertGap)$);
	\node[above=6pt of DoveB,anchor=base] {Dove};
	\node[above=6pt of HawkB,anchor=base] {Hawk};
	\node[left=0.8cm of DoveA,anchor=center] {Dove};
	\node[left=0.8cm of HawkA,anchor=center] {Hawk};
	\node[left=2cm of midLeft,rotate=90,align=center,anchor=center] (AnimalA) {Animal $A$};
	\node[above=18pt of topMid] (AnimalB) {Animal $B$};
	
	\end{tikzpicture}
\end{center}
	\caption{Payoffs to animals $A$ and $B$ in one stage game of our modified HD model. The value of the resource is $1$. When both animals play Hawk, there is a fight that is resolved probabilistically, such that the chance of animal $A$ winning the fight is $p_A$. The winner obtains a payoff of $1 - c_W$ (the value of the resource minus its cost of fighting),
	 and the loser obtains $-c_L$ (a fixed penalty for losing).}
	\label{F:ModifiedHD}
\end{figure}

Our modified HD model has three dimensionless parameters: $c_W$, $c_L$, and $p_A$.
We assume that $c_W$ and $c_L$ are fixed for a given species and that both animals $A$ and $B$ know the values of $c_W$ and $c_L$. By contrast, $p_A$ depends on the different fighting abilities (as measured by RHP) of animals $A$ and $B$.

Let $R_A$ and $R_B$, respectively, denote the RHPs of animals $A$ and $B$. The probability that $A$ wins a fight is $p_A = \varphi(R_A, R_B)$, 
where $\varphi(x,y)$ is some function that satisifies $0 \leq \varphi \leq 1$. We require that $\varphi_x(x,y) > 0$, as an increased RHP implies a higher probability of winning a fight, and that $\varphi(x,y) = 1 - \varphi(y,x)$, so the chance of winning a fight is independent of whether we label an animal as $A$ or $B$. When RHPs are given by positive real numbers, one suitable choice is
\begin{equation}
	\varphi(R_A,R_B) = \frac{R_A}{R_A+R_B}\,,
\label{EqDefn:RHPToProb}
\end{equation}
which enables further simplifications in certain scenarios. If we draw the RHPs from an exponential distribution, the corresponding $p_A$ comes from a uniform distribution. More generally, if we draw the RHPs from a gamma distribution with shape parameter $k$ (and any rate parameter), 
$p_A \sim \mathrm{Beta}(k,k)$. In this paper, we draw $p_A$ from such a beta distribution, and we assume that animals $A$ and $B$ have prior knowledge of this distribution, even if they do not have specific knowledge of $p_A$. 


\subsection{Terminology and classification of strategies}
\label{S:Terminology}

We are concerned with comparing different strategies for playing the iterated HD (IHD) game that we defined in Section \ref{S:HawkDove}. 

Each interaction of animals $A$ and $B$ (where they play the `stage game' in Figure \ref{F:ModifiedHD}) is a `round'. In a round, each animal chooses a `move' of either Hawk or Dove. As in \citet{Houston1991} but unlike \citet{MestertonGibbons2014}, Hawk and Dove are the only possible moves in a round. Our model does not include any concept of resource ownership, and we do not consider Bourgeois or anti-Bourgeois strategies.
If both animals choose Hawk as their move in a round, we say that a `fight' occurs.

The overall game, which consists of a large number of rounds, is a `contest' between animals $A$ and $B$. We evaluate the `total payoff' to each animal by summing the discounted payoffs to animals $A$ and $B$ over all rounds \citep{FujiwaraGreveGameTheory}. Taking $\gamma \in (0,1)$ to be the discount rate and $\rho^{(A)}_k$ to be the payoff to animal $A$ in round $j$, the total payoff to $A$ at the end of a contest is 
\begin{equation}
	\varphi_A = \sum_{j=1}^\infty \gamma^{j-1} \rho^{(A)}_j\,.
\end{equation}

An animal's `strategy' is a set of rules that enable it to determine the move that it plays in each round. A strategy specifies a move for the first round and a rule, based on the outcomes of the previous rounds, for determining which move to select in each subsequent round. 
Strategies can be probabilistic, with a probability between $0$ and $1$ of playing Dove in a particular round. Animals $A$ and $B$ may use different strategies; for simplicity, we describe strategies from the perspective of animal $A$.

We distinguish between three types of strategies: `simple', `informed', and `learning'. 
When animal $A$ pursues a simple strategy, it does not use knowledge of $p_A$ to inform its behaviour. 
When animal $A$ pursues an informed strategy, it begins a contest with perfect knowledge of $p_A$ and uses this information to guide its behaviour. Such a strategy is relevant only when animal $A$ is able both to assess its opponent perfectly without any fighting and to exploit that assessment in its choices.
When animal $A$ pursues a learning strategy, it begins with limited information about $p_A$, but it uses information from fights to update its beliefs about $p_A$ and uses its beliefs about $p_A$ to guide its choices. In Table \ref{T:StrategyTypes}, we summarize the three strategy types.

	\begin{table}[ht]
	\centering
	\footnotesize 
\setlength\tabcolsep{5pt}
\begin{tabular}{|c|c|c|c|}
\hline
\multirow{3}[2]{*}{Strategy type} & Uses information & Available to animals  & Available to animals  \bigstrut[t]\\
      &  about $p_A$  to & with perfect knowledge & with no knowledge  \\
      & inform behavior &  of $p_A$ at start of contest & of $p_A$ at start of contest \bigstrut[b]\\
\hline
Simple & No    & Yes   & Yes \bigstrut\\
\hline
Informed & Yes   & Yes   & No \bigstrut\\
\hline
Learning & Yes   & No    & Yes \bigstrut\\
\hline
\end{tabular}%

	\normalsize
	\caption{A summary of the three different strategy types. Because animals that pursue simple strategies do not use $p_A$ to inform their behaviour, simple strategies are always available to animals, even if they begin a contest without knowledge of $p_A$.}	
	\label{T:StrategyTypes}
\end{table}

In our analysis of simple strategies, we consider only `memory-$1$ strategies' \citep{Press2012,Nowak1993}, in which the probability of playing Dove
 in a round depends only on the moves of the two animals in the previous round. As we discuss in more detail in Section \ref{S:InformedStrats}, one can represent simple memory-$1$ strategies using $\substratDoveProbs$, a vector of probabilities \citep{Nowak1993,Press2012}. Our focus on memory-$1$ strategies places some restrictions on the variety of simple strategies that we can consider, but many important strategies --- including Tit for Tat \citep{AxelrodEvolCooperation}, Grim Trigger \citep{Friedman1971} (which we call `Grim'), Pavlov \citep{Nowak1993}, and extortionate strategies \citep{Press2012} --- are examples of memory-$1$ strategies. 

Our analysis of informed and learning strategies also focuses on memory-$1$ strategies. In an informed memory-$1$ strategy, the probability of playing Dove in a round depends both on $p_A$ and on the moves from the previous round. That is, an informed memory-$1$ strategy is one in which an animal uses knowledge of $p_A$ to choose an element from a set of memory-$1$ `substrategies', each of which can be represented in the standard vector format of a simple strategy (see Section \ref{S:InformedStrats}). For a learning memory-$1$ strategy, the probability of playing Dove in a round depends both on an animal's current beliefs about $p_A$ and on the moves from the previous round. That is, the results of all previous fights inform an animal's beliefs about $p_A$, and it uses these beliefs to determine the substrategy that it employs in a round.

An animal needs a learning strategy only if it does not begin with perfect information about $p_A$. An animal with knowledge of $p_A$ is able to use a simple strategy or an informed strategy, whereas an animal without this knowledge can use either a simple strategy or a learning strategy. Determining optimal strategies for animals with knowledge of $p_A$ is the focus of Section \ref{S:Informed}, and determining optimal strategies for animals without knowledge of $p_A$ is the focus of Section \ref{S:Learning}.

Even with our consideration of only memory-$1$ strategies, the set of possible strategies for this game is infinite. Additionally, there are well-established difficulties with assessing the quality of strategies for iterated games. For discount rates that are sufficiently close to $1$, there are `folk theorems' that imply that there is a very large class of strategies that are subgame-perfect Nash equilibria \citep{Friedman1971,FujiwaraGreveGameTheory}.

Our approach for assessing the success of a strategy is analogous to the methods that were proposed in \citet{MaynardSmith1973} to compare strategies for a single multi-stage conflict between two animals. We consider a limited set of plausible strategies and use simulations to evaluate the outcomes of contests for every possible pair of strategies. Following \citet{AxelrodEvolCooperation}, we refer to the set of all contests between strategies as a `tournament'. As in \citet{MaynardSmith1973}, our contests involve probabilistic elements; to account for this, we run multiple simulations for each pair of strategies and report the mean payoff to each animal. We then present these results in a matrix, from which we evaluate which strategies are evolutionarily stable.

An `evolutionarily stable strategy' (ESS) is based on the following idea: if all members of a population adopt an ESS, a mutant strategy cannot successfully invade the population \citep{MaynardSmithEvolution}. In mathematical terms, let $E(T,S)$ represent the expected payoff to animal $A$ when it pursues strategy $T$ and animal $B$ pursues strategy $S$. A strategy $S$ is an ESS if $E(T,S) \leq E(S,S)$ for all strategies $T \neq S$, with the additional condition that $E(S,T) > E(T,T)$ whenever $E(T,S) = E(S,S)$. Not all games have ESSs, some games have multiple ESSs, and sometimes probabilistic combinations of strategies can be ESSs; these are called `mixed ESSs'. \citep{MaynardSmith1973,FujiwaraGreveGameTheory}.
In many cases, it is possible to use the matrix of payoffs from a tournament to compute the set of all possible ESSs \citep{Haigh1975,BroomGameTheoreticalBiology}.


\subsection{A Bayesian approach to learning}
\label{S:IntroToLearning}

An animal that pursues a learning strategy begins a contest with complete knowledge of $c_W$ and $c_L$, but with no information about $p_A$ beyond the fact that $p_A \sim \mathrm{Beta}(k,k)$ for a specified $k$. The animal uses information from its interactions with its opponent to improve its estimate of $p_A$. Various methods have been developed for implementing learning in repeated games with incomplete information \citep{AumannMaschlerRepeatedGames,Sastry1994,Watkins1992,Jordan1995}. Many of these approaches are very general and sophisticated, but they depend on the assumption that each player is rational and can therefore exploit the assumed rationality of their opponent to obtain information. 

In the present work, by contrast, we use a Bayesian approach to incorporate the information obtained from fights into an animal's beliefs about $p_A$, and we assume that its beliefs about $p_A$ do not change if there is no fight. We assume that $p_A$ is constant in time, so we can treat fights as independent Bernoulli trials with a fixed probability of success. Using Bayes' rule, we update estimates of $p_A$ as an animal acquires information from its fights \citep{GelmanBayesian}. The beta distribution is a conjugate prior of the Bernoulli distribution, so if one can represent an animal's initial beliefs about its probability of winning a fight using a beta distribution, then these beliefs remain a beta distribution (but with altered parameter values) as it obtains information from its fights \citep{GelmanBayesian,Trimmer2011}.

There has been some discussion in the biological literature about whether animals (including humans) are capable of Bayesian updating \citep{Trimmer2011,Valone2006,McNamara2006}. The consensus of such work is that observations of vertebrate behaviour (especially with foraging) are consistent with Bayesian updating. The main challenges of taking a Bayesian approach are determining the parameters for the prior distribution and deciding how animals should use their current estimate of the distribution of $p_A$ to inform their behaviour. In Section \ref{S:Learning}, we propose and compare a variety of learning strategies that uninformed animals can use to assess their probability of winning a fight, and we discuss how this information influences which strategy they adopt.


\section[Analysis of strategies for animals with knowledge of pA]{Analysis of strategies for animals with knowledge of $p_A$}
\label{S:Informed}

When cost-free or low-cost observations are reliable indicators of RHPs, it is reasonable to assume that all animals begin contests with knowledge of $p_A$. In this section, we seek ESSs among the strategies that are available to an animal with knowledge of $p_A$. Finding optimal informed strategies will also guide the development of plausible learning strategies in Section \ref{S:Learning}.


\subsection{Stage-game classification and its implications}
\label{S:StageGameClassification}

Suppose that two animals are playing against each other in the modified HD game of Figure \ref{F:ModifiedHD} and that they both know $p_A$. Replacing the random outcome of a fight with the expected value of the interaction, we obtain the game in Figure \ref{F:AveragedHD}, where the expected payoffs $\mu_A$ and $\mu_B$ to animals $A$ and $B$ when they both play Hawk are
\begin{align}
\label{muA-Defn}
	\mu_A &= -c_L + p_A (1 - c_W + c_L)\,, \\
\label{muB-Defn}
	\mu_B &= 1 - c_W - p_A(1 - c_W + c_L)\,.
\end{align}

\begin{figure}
\begin{center}
	\tikzsetnextfilename{HawkDoveAveraged}
	\begin{tikzpicture}
	
	\coordinate (halfVertGap) at (0,0.7);
	\coordinate (halfHorGap) at (1.4,0);
	\node[align=center] (HDPayoff) at (0,0) {$(1,0)$};
	\node[align=center] (HHPayoff) at ($(halfHorGap)+(halfHorGap)$) {$(\mu_A,\mu_B)$};
	\node[align=center] (DDPayoff) at ($(halfVertGap)+(halfVertGap)$) {$(\tfrac{1}{2},\tfrac{1}{2})$};
	\node[align=center] (DHPayoff) at ($(HHPayoff) + (DDPayoff)$) {$(\tfrac{1}{2},\tfrac{1}{2})$};
	\coordinate (topLeft) at  ($(DDPayoff) - (halfHorGap) + (halfVertGap)$);
	\coordinate (topRight) at ($(DHPayoff) + (halfHorGap) + (halfVertGap)$);
	\coordinate (btmLeft) at  ($(HDPayoff) - (halfHorGap) - (halfVertGap)$);
	\coordinate (btmRight) at ($(HHPayoff) + (halfHorGap) - (halfVertGap)$);
	\coordinate (topMid) at   ($(DDPayoff) + (halfHorGap) + (halfVertGap)$);
	\coordinate (btmMid) at   ($(HDPayoff) + (halfHorGap) - (halfVertGap)$);
	\coordinate (midLeft) at  ($(HDPayoff) - (halfHorGap) + (halfVertGap)$);
	\coordinate (midRight) at ($(HHPayoff) + (halfHorGap) + (halfVertGap)$);
	\draw (topLeft) rectangle (btmRight);
	\draw (topMid) -- (btmMid);	
	\draw (midLeft) -- (midRight);
	
	\coordinate (DoveA) at ($(DDPayoff) - (halfHorGap)$);
	\coordinate (DoveB) at ($(DDPayoff) + (halfVertGap)$);
	\coordinate (HawkA) at ($(HDPayoff) - (halfHorGap)$);
	\coordinate (HawkB) at ($(DHPayoff) + (halfVertGap)$);
	\node[above=6pt of DoveB,anchor=base] {Dove};
	\node[above=6pt of HawkB,anchor=base] {Hawk};
	\node[left=0.8cm of DoveA,anchor=center] {Dove};
	\node[left=0.8cm of HawkA,anchor=center] {Hawk};
	\node[left=2cm of midLeft,rotate=90,align=center,anchor=center] (AnimalA) {Animal $A$};
	\node[above=18pt of topMid] (AnimalB) {Animal $B$};
	
	\end{tikzpicture}
	\end{center}
	\caption{Expected values of the payoffs to animals $A$ and $B$ in one 
	round of our modified HD game.
		We give the expected payoffs from fighting, $\mu_A$ and $\mu_B$ in equations \eqref{muA-Defn} and \eqref{muB-Defn}.
		The character of the game (and the corresponding best strategies) depend on the values of $\mu_A$ and $\mu_B$.}
	\label{F:AveragedHD}
\end{figure}

The optimal strategies for informed animals $A$ and $B$ in an iterated game depend on the ordering of the payoffs of the different interactions in the stage game; these, in turn, depend on the values of $\mu_A$ and $\mu_B$. The stage-game payoffs associated with Dove--Dove and Hawk--Dove plays by the animals are $(\tfrac{1}{2},\tfrac{1}{2})$ and $(1,0)$, respectively. We can thus classify the different stage games according to the values of $\mu_A$ and $\mu_B$ relative to each other and to $0$, $\tfrac{1}{2}$, and $1$. Because $\mu_A < 1$, $\mu_B < 1$, and $\mu_A + \mu_B < 1$, there are $10$ possible orderings of the payoffs. Assuming without loss of generality that $\mu_A > \mu_B$, we enumerate the $5$ remaining orderings in Table \ref{T:OrdinalGames}. We obtain the other $5$ orderings by swapping $\mu_A$ and $\mu_B$.

As one can see in Table \ref{T:OrdinalGames}, we name the classifications of the stage game based on the ordering of payoffs to animals $A$ and $B$. 
For example, if animal $A$ is faced with a `Snowdrift' situation and animal $B$ is faced with a `Deadlock' situation, we have a `Snowdrift--Deadlock' game.

All possible stage games other than the Snowdrift game have a unique Nash equilibrium. In the Snowdrift game, for which $\mu_B$ and $\mu_A$ are both negative, there are three Nash equilibria: (Hawk, Dove), (Dove, Hawk), and a mixed Nash equilibrium in which animal $A$ plays Dove with probability $2|\mu_A|/(1 + 2|\mu_A|)$ and animal $B$ plays Dove with probability $2|\mu_B|/(1 + 2|\mu_B|)$.

\begin{table}[ht]
	\centering 
	\begin{tabular}{|c| c| c|} 
		\hline 
		Ordering & Name & Nash Equilibria \\ [0.5ex] 
		\hline 
		$\mu_B < \mu_A < 0 < \tfrac{1}{2} < 1$ & Snowdrift ($A$ beats $B$) & (D,H), (H,D), or mixed  \\
		$\mu_B < 0 < \mu_A < \tfrac{1}{2} < 1$ & PD--Snowdrift & (H,D) \\
		$\mu_B < 0  < \tfrac{1}{2} < \mu_A < 1$ & Deadlock--Snowdrift & (H,D) \\
		$0 < \mu_B < \mu_A  < \tfrac{1}{2} < 1$ & PD ($A$ beats $B$) & (H,H) \\
		$0 < \mu_B  < \tfrac{1}{2} < \mu_A  < 1$ & Deadlock--PD & (H,H) \\
		[1ex] 
		\hline 
	\end{tabular}
	\caption{Classification of the stage game according to the values of $\mu_A$ and $\mu_B$, with $\mu_A > \mu_B$.
		We base the nomenclature on the ordering of payoffs to 
		the animals. For example, a `PD--Snowdrift' game is one in which the ordering of payoffs to animal $A$ is identical to that in the classical Prisoner's Dilemma game, whereas the ordering of payoffs to animal $B$ is identical to that in the classical Snowdrift game. 
		} 
	\label{T:OrdinalGames}
\end{table}

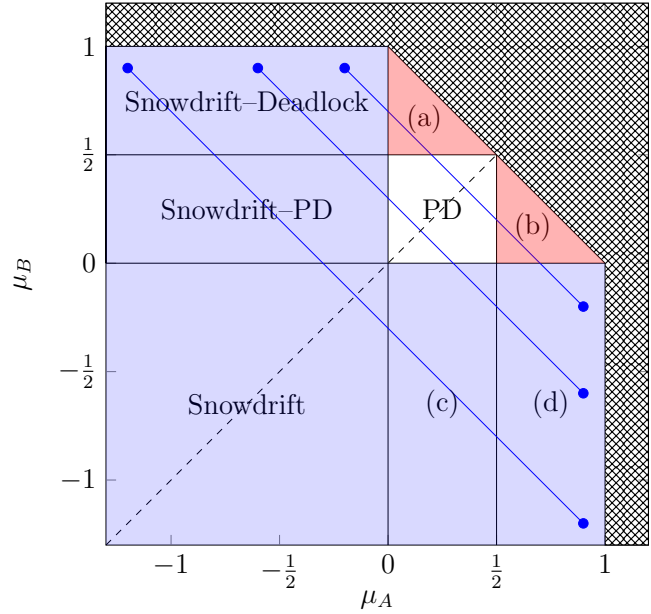
\begin{figure}[htp]
	\centering
	\tikzsetnextfilename{muValuesAndClassification}
	\begin{tikzpicture}[scale=0.9]
	\begin{axis}[
	ymin=-1.3,ymax=1.2,xmin=-1.3,xmax=1.2,
	xlabel={$\mu_A$},
	ylabel={$\mu_B$},
	every inner x axis line/.append style={<->},
	every inner y axis line/.append style={<->},
	xtick={0},
	ytick={0},
	extra x ticks={-1,-0.5,0.5,1},
	extra y ticks={-1,-0.5,0.5,1},
	extra x tick labels={$-1$,$-\tfrac{1}{2}$,$\tfrac{1}{2}$, $1$}, 
	extra y tick labels={$-1$,$-\tfrac{1}{2}$,$\tfrac{1}{2}$, $1$}, 
	width={0.7\textwidth},
	height={0.7\textwidth}
	]
	\addplot[color=black] coordinates{(0.5,-1.3) (0.5,0.5)};
	\addplot[color=black] coordinates{(-1.3,0.5) (0.5,0.5)};
	\addplot[color=black] coordinates{(0,-1.3)   (0,1)};
	\addplot[color=black] coordinates{(-1.3,0)   (1,0)};
	\addplot[color=blue, mark=*] coordinates{(-0.2,0.9) (0.9,-0.2)};
	\addplot[color=blue, mark=*] coordinates{(-0.6,0.9) (0.9,-0.6)};
	\addplot[color=blue, mark=*] coordinates{(-1.2,0.9)   (0.9,-1.2)};
	\addplot[color=black, dashed] coordinates{(-1.3,-1.3) (0.5,0.5)};
	\node at (axis cs:-0.65, 0.75)   {Snowdrift--Deadlock};
	\node at (axis cs:-0.65, 0.25)   {Snowdrift--PD};
	\node at (axis cs:-0.65, -0.65)   {Snowdrift};
	\node at (axis cs:0.25, 0.25)   {PD};
	\node at (axis cs:1/6,  2/3)    {(a)};
	\node at (axis cs:2/3,  1/6)    {(b)};
	\node at (axis cs:0.25, -0.65)   {(c)};
	\node at (axis cs:0.75, -0.65)   {(d)};
	\addplot[pattern=crosshatch] coordinates{(-1.3,1) (0,1) (1,0) (1,-1.3) (1.2,-1.3) (1.2,1.2) (-1.3,1.2)} \closedcycle;
	\addplot[draw opacity=0, fill opacity=0.3, fill=red] coordinates{(0.5,0) (1,0) (0,1) (0,0.5) (0.5,0.5) (0.5,0)} \closedcycle;
	\addplot[draw opacity=0, fill opacity=0.15, fill=blue] coordinates{(-1.3,-1.3) (1,-1.3) (1,0) (0,0) (0,1) (-1.3,1) (-1.3,-1.3)} \closedcycle;
	\end{axis}
	\end{tikzpicture}
	\caption{
	The classification of the stage game depends on $\mu_A$ and $\mu_B$, which are the expected payoffs of 
	fighting for animals $A$ and $B$, respectively.
		We mark the case with $p_A = 0.5$ (in which animals $A$ and $B$ are evenly matched) with a dashed line, and we separate the different stage-game classifications with solid lines.
		We shade the region that is inaccessible when $c_L$ and $c_W$ are positive.
		In addition to the stage-game classifications that we label in the diagram, (a) designates a PD--Deadlock game, (b) designates Deadlock--PD, (c) designates PD--Snowdrift, and (d) designates Deadlock--Snowdrift.
		The blue lines show the possible values of
		$\mu_A$ and $\mu_B$ from
		varying $p_A$ for three different choices of $c_W$ and $c_L$.
		From left to right, the blue lines correspond to the parameter pairs $(c_W,c_L) = (0.1,1.2)$, $(c_W,c_L) = (0.1,0.6)$, and $(c_W,c_L) = (0.1,0.2)$. In Section \ref{S:InformedDiscussion}, we discuss the behaviour of two informed animals that both pursue the optimal strategy from
		Section \ref{S:InformedTournament}. 
		This behaviour depends on the stage-game classification; we illustrate this dependence by coluring regions of the figure.
		We use pink when informed animals who pursue an optimal strategy play Hawk against each other in all but the first round, white when they play Dove against each other in every round, and light blue when the animal with a larger RHP plays Hawk in every round and the animal with a smaller RHP plays Dove in every round.}
	\label{F:muValuesAndClassification}
\end{figure}

The different classifications of the stage game correspond to different regions in ($\mu_A$, $\mu_B$)-space (see Figure \ref{F:muValuesAndClassification}). For fixed $c_W$ and $c_L$, changing $p_A$ leads to changes in $\mu_A$ and $\mu_B$, which in turn correspond to changes in the classification of the stage game. A fixed choice of $c_W$ and $c_L$ defines a unique line segment in Figure \ref{F:muValuesAndClassification} that connects $(-c_L,1-c_W)$ to $(1-c_W,-c_L)$. The center of each line segment occurs when $\mu_A= \mu_B = \frac{1}{2}(1 - c_W - c_L)$, for which $p_A = \frac{1}{2}$. The blue lines in Figure \ref{F:muValuesAndClassification} show three examples for different values of the parameters $c_W$ and $c_L$.
As one increases $p_A$, one moves along such a line in Figure \ref{F:muValuesAndClassification} from the top left to the bottom right.

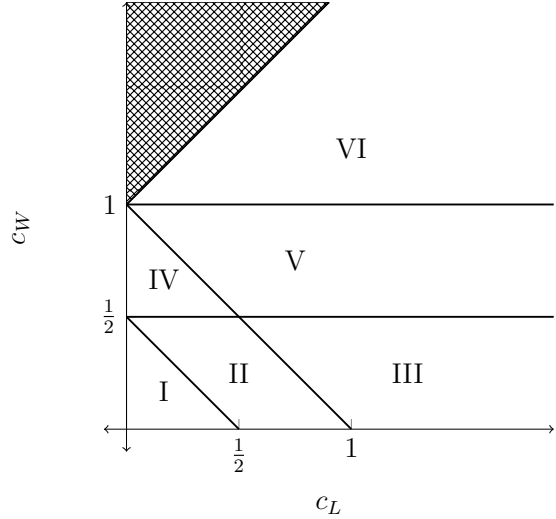
\begin{figure}[htp]
	\centering
	\tikzsetnextfilename{StageGameRegimes}
	\begin{tikzpicture}[scale=0.9]
	\begin{axis}[
	axis x line*=center,
	axis y line*=center,
	ymin=-0.1,ymax=1.9,xmin=-0.1,xmax=1.9,
	xlabel={$c_L$},
	ylabel={$c_W$},
	every inner x axis line/.append style={<->},
	every inner y axis line/.append style={<->},
	xtick={0},
	ytick={0},
	extra x ticks={0.5,1},
	extra y ticks={0.5,1},
	extra x tick labels={$\tfrac{1}{2}$, $1$}, 
	extra y tick labels={$\tfrac{1}{2}$, $1$}, 
	width={0.6\textwidth},
	height={0.6\textwidth}
	]
	\addplot[color=black, thick] coordinates{(0,0.5) (0.5,0)};
	\addplot[color=black, thick] coordinates{(0,0.5) (1.9,0.5)};
	\addplot[color=black, thick] coordinates{(0,1)   (1,0)};
	\addplot[color=black, thick] coordinates{(0,1)   (1.9,1)};
	\addplot[color=black, thick] coordinates{(0,1)   (0.9,1.9)};
	\node at (axis cs:1/6,  1/6)   {I};
	\node at (axis cs:0.5,  0.25)  {II};
	\node at (axis cs:1.25, 0.25)  {III};
	\node at (axis cs:1/6,  2/3)   {IV};
	\node at (axis cs:0.75, 0.75)  {V};
	\node at (axis cs:1,    1.25)  {VI};
	\addplot[pattern=crosshatch] coordinates{(0,1) (0.9,1.9) (0,1.9)} \closedcycle;
	\end{axis}
	\end{tikzpicture}
	\caption{As $p_A$ changes, the classification of the stage game changes.
		We show the different regimes in ($c_W$,$c_L$)-space that lead to different sequences of stage-game classifications. We shade the region that is inaccessible due to the requirement that $1-c_W > -c_L$.
		We outline the stage games 
		in Table \ref{T:StageGameRegimes}, and one can also see them in Figure \ref{F:muValuesAndClassification}.}
	\label{F:StageGameRegimes}
\end{figure}

As we can see from the example line segments in Figure \ref{F:muValuesAndClassification}, the sequence of stage games as $p_A$ increases from $0$ to $1$ depends on the values of $c_W$ and $c_L$. Considering all allowable possibilities in $(c_W,c_L)$-space (see Section \ref{S:HawkDove}), there are six different parameter regimes, which are associated with different sequences of stage-game classifications. We show the parameter regimes in Figure \ref{F:StageGameRegimes}, and we detail the associated sequences of stage-game classifications in Table \ref{T:StageGameRegimes}.

Parameter regimes IV, V, and VI encompass a narrower range of stage games than parameter regimes I, II, and III.
For example, we see from Table \ref{T:StageGameRegimes} that the stage-game classifications that occur in parameter regime IV as we change $p_A$ 
from $0$ to $\tfrac{1}{2}$ are a subset of those that occur in parameter regime II for the same values of $p_A$. For example, the transition from Snowdrift--Deadlock to Snowdrift--PD at $p_A  = (\tfrac{1}{2} - c_W)/q$ does not occur in parameter regime IV, as $(\tfrac{1}{2} - c_W)/q < 0$ in this case.
Similarly, the stage-game classifications that occur for $p_A \in [0,1/2]$ in parameter regime V are a subset of those that occur for $p_A \in [0,1/2]$ in parameter regime III; and the stage-game classifications for $p_A \in [0,1/2]$ in parameter regime VI are a subset of those for $p_A \in [0,1/2]$ in parameter regime V (and therefore also a subset of those for $p_A \in [0,1/2]$ in parameter regime III).

To focus on cases in which optimal strategy can vary significantly as $p_A$ changes, we concentrate our analysis on parameter regimes I, II, and III.

\begin{table}[ht]
	\centering 
	\begin{tabular}{|c| c| c|} 
		\hline 
		Regime & Range of $p_A$ values & Classification \\ [0.5ex] 
		\hline 
		I   & $0 < p_A < c_L / q$                           & Snowdrift--Deadlock  \\
		& $ c_L / q < p_A < (\tfrac{1}{2} - c_W)/q$     & PD--Deadlock \\
		& $(\tfrac{1}{2} - c_W)/q < p_A < \tfrac{1}{2}$ & PD ($B$ beats $A$) \\  
		[1ex] \hline
		II  & $0 < p_A < (\tfrac{1}{2} - c_W)/q$            & Snowdrift--Deadlock  \\
		& $(\tfrac{1}{2} - c_W)/q < p_A < c_L / q$      & Snowdrift--PD  \\
		& $c_L / q < p_A < \tfrac{1}{2}$                & PD ($B$ beats $A$)  \\
		[1ex] \hline 
		III & $0 < p_A < (\tfrac{1}{2} - c_W)/q$            & Snowdrift--Deadlock  \\
		& $(\tfrac{1}{2} - c_W)/q < p_A < (1 - c_W)/q$  & Snowdrift--PD  \\
		& $(1 - c_W)/q < p_A < \tfrac{1}{2}$            & Snowdrift ($B$ beats $A$)  \\
		[1ex] \hline 
		IV  & $0 < p_A < c_L/q$            & Snowdrift--PD  \\
		& $c_L/q < p_A < \tfrac{1}{2}$ & PD ($B$ beats $A$) \\
		[1ex] \hline 
		V   & $0 < p_A < (1 - c_W)/q$                       & Snowdrift--PD \\
		& $(1 - c_W)/q < p_A < \tfrac{1}{2}$            & Snowdrift ($B$ beats $A$)  \\
		[1ex] \hline     
		VI  & $0  < p_A < \tfrac{1}{2}$                     & Snowdrift ($B$ beats $A$)  \\
		[1ex] \hline     
	\end{tabular}
	\caption{Changes to the classification of the stage game as $p_A$ increases from $0$ to $\tfrac{1}{2}$ in each of the parameter regimes 
		from Figure \ref{F:StageGameRegimes}.
		For convenience, we define the notation 
		$q = 1 - c_W + c_L$.
		We obtain the classifications of the stage game for $p_A > \frac{1}{2}$ by symmetry. For $p_A = \frac{1}{2}$, animals $A$ and $B$ are evenly matched, and the game is a classical PD
		(in regimes I, II, and IV) or Snowdrift game (in regimes III, V, and VI), where both animals have
		the same expected payoff from a fight.
		In our numerical experiments, we concentrate on examples with $c_W < \tfrac{1}{2}$, so we are in one of regimes I--III.
	} 
	\label{T:StageGameRegimes}
\end{table}


\subsection{Defining informed strategies}
\label{S:InformedStrats}

We consider seven informed strategies: `Bully', `Nash', `Mixed Nash', `Selfish', `Snowdrift TfT', `PD TfT', and `Informed TfT'. In Table \ref{T:PlausibleStrategies}, we describe the substrategies that are pursued by these informed strategies for different stage games. Supposing that animal $A$ is the one that is playing the strategy and that animal $B$ is its opponent, we summarize the seven informed strategies as follows:
\begin{itemize}
	\item{`Bully' always plays Hawk if $p_A > \tfrac{1}{2}$ (i.e., when it is more likely than not to beat its opponent) and always plays Dove if $p_A < \tfrac{1}{2}$.}
	\item{`Nash' pursues a Nash-equilibrium substrategy for each move. 
	In a Snowdrift game, `Nash' uses the difference in RHP to decide which pure Nash equilibrium to pursue; it always plays Hawk if $p_A > \tfrac{1}{2}$ and always plays Dove if $p_A < \tfrac{1}{2}$.}
	\item{`Mixed Nash' pursues a Nash-equilibrium substrategy for each move. 
	In a Snowdrift game, `Mixed Nash' pursues a mixed substrategy in which its probability of playing Dove is $2|\mu_A|/(1 + 2|\mu_A|)$.}
	\item{`Selfish' chooses a substrategy based on its expected payoff from a fight. 
	It always plays Hawk if $\mu_A > \tfrac{1}{2}$, pursues a Tit for Tat substrategy if $0 < \mu_A < \tfrac{1}{2}$, and always plays Dove if $\mu_A < 0$.}
	\item{`Snowdrift TfT' pursues a Tit for Tat substrategy when the stage game is Snowdrift or Snowdrift--PD, but it otherwise follows the Nash equilibrium.}
	\item{`PD TfT' pursues a Tit for Tat substrategy when the stage game is PD or PD--Deadlock, but it otherwise follows the Nash equilibrium.}
	\item{`Informed TfT' pursues a Tit for Tat substrategy, except when this is clearly unsuitable. Specifically, 
	it always plays Hawk in a Deadlock--Snowdrift or Deadlock--PD stage game and always plays Dove in a Snowdrift--Deadlock stage game, but it otherwise pursues Tit for Tat.}
\end{itemize}

\begin{table}[ht]
	\centering 
	\begin{tabular}{|c| c c c c c c c|}
		\hline
		Stage game            & Bul & N  & MN  & Slf & SDT & PDT & InT 
		\\ [0.5ex] 
		\hline
		Deadlock--Snowdrift   & H   & H   & H   & H   & H   & H   & H
		\\
		Deadlock--PD          & H   & H   & H   & H   & H   & H   & H 
		\\
		PD--Snowdrift         & H   & H   & H   & TfT & H   & H   & TfT
		\\
		PD ($A$ beats $B$)        & H   & H   & H   & TfT & H   & TfT & TfT
		\\
		PD ($B$ beats $A$)        & D   & H   & H   & TfT & H   & TfT & TfT
		\\
		PD--Deadlock          & D   & H   & H   & TfT & H   & TfT & TfT
		\\
		Snowdrift ($A$ beats $B$) & H   & H   & Mix & D   & TfT & H   & TfT
		\\
		Snowdrift ($B$ beats $A$) & D   & D   & Mix & D   & TfT & D   & TfT
		\\
		Snowdrift--PD         & D   & D   & D   & D   & TfT & D   & TfT
		\\
		Snowdrift--Deadlock   & D   & D   & D   & D   & D   & D   & D
		\\
		[1ex] 
		\hline 
	\end{tabular}
	\caption{%
		Summary of substrategies when animal $A$ play each of our informed strategies: Bully (`Bul'), Nash (`N'), Mixed Nash (`MN'), Selfish (`Slf'), Snowdrift TfT (`SDT'), PD TfT (`PDT'), and Informed TfT (`InT'). 
		The entries in the table indicate the substrategy that is pursued by animal $A$ when the stage-game classification is the one in the left column. 
		The strategies in the table are Hawk (`H'), Dove (`D'), Tit for Tat (`TfT'), and mixed substrategies (`Mix'). 
		See the text for details.
	} 
	\label{T:PlausibleStrategies}
\end{table}

We do not specify substrategies for the non-generic games that occur when $p_A$ is at a `critical' value at a boundary between stage-game classifications (see Table \ref{T:StageGameRegimes}), because the probability that this occurs in a simulation is vanishingly small. If such a critical situation occurs, we assume that each animal chooses the most aggressive of its available strategies. That is, each animal chooses the strategy that it would pursue if its RHP were larger by an arbitrarily small amount for a fixed value of its opponent's RHP. Our software also returns a warning in such a critical case; no such warning occurred in any of our tournaments.

We now discuss our motivation behind the above strategies. 
One approach that animals can take is to pursue a Nash-equilibrium strategy of the stage game in each round. This involves a small complication for the Snowdrift game (where there are three Nash equilibria), but otherwise each animal has a unique strategy in each round. Pursuing the stage-game Nash equilibrium in each round of an iterated game is not necessarily optimal. See \citet{Axelrod1981} for the most famous example, which arises in the IPD. Nevertheless, cooperative strategies, such as Tit for Tat and Grim (in which each animal plays Dove until the opposing animal has played Hawk), can still be very successful and are evolutionarily stable (except to invasion from other cooperative strategies) for sufficiently large values of the discount parameter $\gamma$ \citep{Axelrod1981,FujiwaraGreveGameTheory}. 

We examine informed strategies in which animals frequently follow a Nash-equilibrium strategy, but they may deviate from this for certain stage-game classifications (for example, by playing Tit for Tat when faced with a PD stage game).
Not all deviations from a Nash-equilibrium strategy are appropriate, as a Nash-equilibrium strategy is clearly optimal in certain stage-game classifications.
For example, if animal $A$ is playing a Deadlock--PD or Deadlock--Snowdrift stage game, its best strategy is to play Hawk, regardless of whether animal $B$ plays Hawk or Dove. Consequently, all of our informed strategies pursue an Always Hawk substrategy when faced with a Deadlock--PD or Deadlock--Snowdrift stage game. Similarly, all of our informed strategies pursue an Always Dove substrategy when faced with a Snowdrift--Deadlock stage game.

For other stage-game classifications, the choice of optimal strategy is less obvious.
For simplicity, we assume for each type of stage game that an informed animal pursues one of four different substrategies: (1) always play Hawk, (2) always play Dove, (3) play Tit for Tat, or (4) play a mixed Hawk--Dove strategy. For a given type of stage game, only a strict subset of these strategies is a reasonable choice for a rational animal. For example, the mixed Hawk--Dove strategy can be a Nash equilibrium only for a Snowdrift game, and Always Dove is not a rational option (compared to Always Hawk or Tit for Tat) for an animal in a PD--Snowdrift game, because the lack of punishment \citep[in the sense described by][]{Abreu1988} implies that `Always Dove' cannot be a subgame-perfect equilibrium.


\subsection{Tournaments for comparing informed strategies}
\label{S:InformedTournament}

We use tournaments, which consist of many contests between different pairs of strategies, to compare the performance of our seven informed strategies against each other and against nine 
simple memory-$1$ strategies. Animals that pursue simple strategies do not take advantage of the information that is available from $p_A$. (See Section \ref{S:Terminology} for our definition of `simple'.)
We represent each simple memory-$1$ strategy using $\substratDoveProbs$, a vector of six probabilities. The first five elements give the probabilities of an animal playing Dove when the previous moves by it and its opponent are Dove--Dove, Dove--Hawk, Hawk--Dove, Hawk--Hawk (and the animal loses), and Hawk--Hawk (and the animal wins). The sixth element gives the probability that the animal plays Dove on its first move. The vector $\substratDoveProbs$ is analogous to the strategy vectors in \citet{Nowak1993} and \citet{Press2012}. 

We consider the following simple memory-$1$ strategies: Always Dove, 25\% Dove, 50\% Dove, 75\% Dove, Always Hawk, Tit for Tat \citep{Axelrod1981}, Grim \citep{Friedman1971}, Pavlov \citep{Nowak1993}, and Modified Pavlov.
In Table \ref{T:Memory1Strategies}, we summarize the probabilities of playing Dove (the components of $\substratDoveProbs$) 
that are associated with these strategies. 

An animal that uses the Pavlov strategy plays Dove with very high probability (which we take to be $0.99$ in our tournaments) in response to a previous (Dove, Dove) or (Hawk, Hawk) round; otherwise, it plays Hawk. It plays Hawk with very high probability (which we again take to be $0.99$ in practice) in response to a previous (Hawk, Dove) or (Dove, Hawk) round. Modified Pavlov follows the same `win--stay, lose--shift' philosophy of Pavlov \citep[see][]{Nowak1993}, but instead of treating `Hawk--Hawk' as a loss in all circumstances (as in the original Pavlov), Modified Pavlov treats `Hawk--Hawk' as a loss only if the associated fight is a loss. We assume that an animal that plays Pavlov or Modified Pavlov is equally likely to play Hawk or Dove in its first move. In tournaments that we do not discuss in this paper, we examined versions of Pavlov and Modified Pavlov in which the first move is either definitely Hawk or definitely Dove. These other choices for the first move had little effect on the overall payoffs.

\begin{landscape}
	\begin{table}[ht]
		\centering
\begin{tabular}{|c|c|c|c|c|c|c|}
	\hline
	& \multicolumn{6}{c|}{Probability of playing Dove (otherwise playing Hawk)} \bigstrut\\
	\hline
	Strategy name & After DD & After DH & After HD & After HH (loss) & After HH (win) & On first move \bigstrut\\
	\hline
	Always Hawk & 0     & 0     & 0     & 0     & 0     & 0 \bigstrut\\
	\hline
	25\% Dove & 0.25  & 0.25  & 0.25  & 0.25  & 0.25  & 0.25 \bigstrut\\
	\hline
	50\% Dove & 0.5   & 0.5   & 0.5   & 0.5   & 0.5   & 0.5 \bigstrut\\
	\hline
	75\% Dove & 0.75  & 0.75  & 0.75  & 0.75  & 0.75  & 0.75 \bigstrut\\
	\hline
	Always Dove & 1     & 1     & 1     & 1     & 1     & 1 \bigstrut\\
	\hline
	Tit for Tat & 1     & 0     & 1     & 0     & 0     & 1 \bigstrut\\
	\hline
	Grim  & 1     & 0     & 0     & 0     & 0     & 1 \bigstrut\\
	\hline
	Pavlov & 0.99  & 0.01  & 0.01  & 0.99  & 0.99  & 0.5 \bigstrut\\
	\hline
	Modified Pavlov & 0.99  & 0.01  & 0.01  & 0.99  & 0.01  & 0.5 \bigstrut\\
	\hline
\end{tabular}%

		\caption{Our memory-$1$ strategies.
		Each row gives the components of $\substratDoveProbs$ of
		the specified memory-$1$ strategy.
	    We show the probabilities that an animal plays Dove (otherwise, it plays Hawk) after different move pairs from it and its opponent (listed in that order) in the previous round. We abbreviate Dove as `D' and Hawk as `H', so the numbers in the `After DH' column give the probability of playing Dove when the animal played Dove and its opponent played Hawk in the previous round. The final column gives the probability that an animal plays Dove in the first round.
		}
		\label{T:Memory1Strategies}
	\end{table}
\end{landscape}

In each tournament, we specify the parameters for the cost of fighting ($c_W$ and $c_L$), the discount rate ($\gamma$), and the shape parameter of the symmetric beta distribution of win probabilities ($k$) at the outset, and we use the same values throughout the tournament. In all of our tournaments, we take $c_W = 0.1$, $\gamma = 0.995$, and $k = 1$. In this section, we describe the results of tournaments with three different values of $c_L$ ($0.2$, $0.6$, and $1.2$). Combined with taking $c_W = 0.1$, these values of $c_L$ correspond to different parameter regimes (see Section \ref{S:StageGameClassification} and Figure \ref{F:muValuesAndClassification}).
Specifically, $(c_W,c_L) = (0.1,0.2)$ corresponds to regime I, $(c_W,c_L) = (0.1,0.6)$ corresponds to regime II, and $(c_W,c_L) = (0.1,1.2)$ corresponds to regime III. Our code (in {\sc Matlab}) for running our tournaments is available at \url{https://bitbucket.org/CameronLHall/dominancesharingassessmentmatlab/}.

We now discuss our parameter choices and the probability distribution from which we select $p_A$ 
(i.e., the probability that animal $A$ wins a fight). Taking $k = 1$ corresponds to choosing values of $p_A$ from a uniform distribution between $0$ and $1$. We also perform numerical simulations (which we do not present in the paper) using the arcsine distribution $p_A \sim \mathrm{Beta}(\tfrac{1}{2},\tfrac{1}{2})$ and using a unimodal beta distribution with a reasonably large variance by taking $p_A \sim \mathrm{Beta}(2,2)$. We obtain qualitatively similar results for all three of these choices.

We choose the discount rate $\gamma$ to be close to $1$ so that the overall payoff from a contest is not dominated by payoffs from the first few rounds. Because of this, we use a large number of rounds ($1000$ of them in our simulations) in each contest to minimize numerical error from the finite nature of the contest. Because we take $\gamma = 0.995$ and the maximum (respectively, minimum) payoff from each round is $1$ (respectively, $-c_L$), the overall payoffs from our simulations are no more than $\frac{\gamma^{1000}}{1 - \gamma} \approx 1.33$ smaller than and no more than $-1.33c_L$ larger
greater than for contests with infinitely many rounds. 

We perform strategy-pair comparisons by assigning one strategy to animal $A$ and the other strategy to animal $B$ and then calculating the mean payoff to animal $A$ from a large number of simulations (using a wide range of values of $p_A$). The resulting mean payoff to animal $A$ approximates the expected payoff to animal $A$ in the given strategy pair.
We use variance-reduction techniques \citep{OwenMonteCarlo} in order to improve the efficiency and accuracy of our estimates of expected payoff; the methods used to compute our results are described in Appendix \ref{S:VarianceReduction}.

With $16$ different strategies (seven informed ones and nine simple ones), we anticipate a tournament to consist of $256$ strategy-pair comparisons. In practice, we can reduce the number of strategy-pair comparisons, because certain informed strategies are identical in some parameter regimes. For example, the `Bully' and `Nash' strategies are identical except when the stage game is PD or PD--Deadlock.
Because neither of these stage games occur in parameter regime III, the `Bully' and `Nash' strategies correspond to identical behaviours in this parameter regime, so we can combine them. In our tabulated results (see Appendix \ref{S:ResultsInformed}), we indicate where we combine informed strategies that are identical in a given parameter regime.

In Appendix \ref{S:ResultsInformed}, we give a full description of our results in the form of three tables (see Tables  \ref{T:RegimeIGridOfOutputs}, \ref{T:RegimeIIGridOfOutputs}, and \ref{T:RegimeIIIGridOfOutputs}) that show the mean payoff to animal $A$ when it pursues the strategy in the row and animal $B$ pursues the strategy in the column. To make ESSs easier to identify, we colour the cells in Tables \ref{T:RegimeIGridOfOutputs}, \ref{T:RegimeIIGridOfOutputs}, and \ref{T:RegimeIIIGridOfOutputs} based on the value of $E(T,S) - E(S,S)$. We show cells as red when $E(T,S) - E(S,S) > 0$ and blue when $E(T,S) - E(S,S) < 0$; deeper shades correspond to larger differences. Therefore, a red cell indicates that the strategy in the row is able to invade the strategy in the column. If a strategy is a strong ESS, all cells in the corresponding column are blue, except for the one on the main diagonal.


\subsection{Discussion of tournament results for informed strategies}
\label{S:InformedDiscussion}

From the tables in Appendix \ref{S:ResultsInformed}, we see that no simple strategy is a pure ESS and that at least one informed strategy is able to invade each simple strategy (except the Pavlov strategy in parameter regime III). The only strategy that is a pure ESS in all parameter regimes is PD Tit for Tat (PD TfT); see column L in Table \ref{T:RegimeIGridOfOutputs}, column N in Table \ref{T:RegimeIIGridOfOutputs}, and column J in Table \ref{T:RegimeIIIGridOfOutputs}.

Interpreting Tables  \ref{T:RegimeIGridOfOutputs}, \ref{T:RegimeIIGridOfOutputs}, and \ref{T:RegimeIIIGridOfOutputs} as payoff matrices for the overall contests, one can use the method of \citet{Haigh1975} to determine whether there are any mixed ESSs.
Using this approach, we find that no mixed ESSs are possible in parameter regimes I or II.
In parameter regime III, there is a mixed ESS in which 87\% of the animals pursue a Pavlov strategy and 13\% of animals pursue a Modified Pavlov strategy. This ESS corresponds to an expected payoff to the animals of $94.3$ (which is less than the expected payoff of $99.3$ of the PD TfT ESS). We concentrate on PD TfT in the rest of our discussion, but it is worth highlighting that a combination of simple strategies can be evolutionarily stable when fighting is very costly.

Consider the case in which animals $A$ and $B$ both pursue PD TfT strategies. 
We use the definition of PD TfT from Table \ref{T:PlausibleStrategies} to describe how the behaviour of these animals depends on the model parameters $c_W$, $c_L$, and $p_A$. We give the details of this analysis in Appendix \ref{S:PairBehaviour}. There are three possible pair behaviours (a dominance relationship, continual fights, and sharing of resources) in parameter regime I, two possible pair behaviours (a dominance relationship and sharing of resources) in parameter regimes II and IV, and only one possible pair behaviour (a dominance relationship) in parameter regimes III, V, and VI. We illustrate this in Figure \ref{F:muValuesAndClassification}, which we colour to indicate the values of $\mu_A$ and $\mu_B$ that are associated with the different pair behaviours.


\section[Analysis of strategies for animals with no knowledge of pA]{Analysis of strategies for animals with no knowledge of $p_A$}
\label{S:Learning}

The pair behaviours from Section \ref{S:InformedDiscussion} occur only if both animals in a contest pursue a PD TfT strategy. Because PD TfT is an informed strategy, the observation of such behaviour depends on the ability of the two animals to perfectly assess each other's RHPs (and hence the value of $p_A$) before a contest begins. However, perfect assessment of RHP is unlikely without paying some cost; animals obtain some information from conventional signals, but there is evolutionary pressure towards costliness of signals and assessment to ensure their accuracy \citep{Dawkins1991,Arnott2009}. We incorporate a costly assessment process into our model by assuming that animals start a contest with no information about $p_A$ but use (potentially costly) Hawk--Hawk fights to obtain information about $p_A$. An animal's beliefs about $p_A$ guide its behaviour.


\subsection{Defining learning strategies}
\label{S:LearningDefns}

We model assessment using learning strategies with Bayesian updating (see Section \ref{S:IntroToLearning}).
Suppose that animal $A$ pursues a learning strategy. At the outset of a contest, we represent 
animal $A$'s initial beliefs about $p_A$ as a beta distribution $\mathrm{Beta}(\alpha_0,\beta_0)$.
We need to specify the values of $\alpha_0$ and $\beta_0$ as part of the definition of the learning strategy.
After each Hawk--Hawk fight, we update animal $A$'s beliefs about $p_A$. If animal $A$ has won $m$ fights and lost $n$ fights since the beginning of a contest, we represent its beliefs about $p_A$ by the beta distribution $\mathrm{Beta}(\alpha_0+m,\beta_0+n)$.

In each round, animal $A$ decides its move using a memory-$1$ substrategy $\substratDoveProbs$ (see Section \ref{S:InformedTournament}) that depends on its current beliefs about $p_A$. We need to specify the relationship between animal $A$'s beliefs (represented by the probability distribution $\mathrm{Beta}(\alpha,\beta)$) about $p_A$ in a given round and the memory-$1$ substrategy $\substratDoveProbs$ that it uses in that round as part of the learning strategy.

We base the relationship between animal $A$'s current beliefs and its current substrategy on the PD Tit for Tat informed strategy.
As an animal becomes more certain about $p_A$, its behaviour should converge to that of an informed animal that uses a PD TfT strategy. Therefore, as $\alpha$ and $\beta$ approach infinity with $\frac{\alpha}{\alpha+\beta} \to p_A$, the substrategy that is associated with $\mathrm{Beta}(\alpha,\beta)$ should converge to the PD TfT strategy for that value of $p_A$.

There are numerous plausible learning strategies that achieve this, and there are numerous possible choices for $\alpha_0$ and $\beta_0$. As with our informed strategies (see Section \ref{S:InformedStrats}), we define a small set of plausible learning strategies and compare them using tournaments. We consider four types of learning strategies:
\begin{enumerate}
	\item[(1)]{weighted learning strategies,}
	\item[(2)]{mean-based learning strategies,}
	\item[(3)]{median-based learning strategies, and}	
	\item[(4)]{high-quantile learning strategies.}
\end{enumerate}
For fixed $c_W$ and $c_L$, recall that the substrategy that is associated with PD TfT changes abruptly as $p_A$ passes through the certain critical values. As we describe in Appendix \ref{S:PairBehaviour}, we define $\tilde{p}_1$ and $\tilde{p}_2$ such that if animal $A$ uses PD TfT, it pursues Always Dove when $0 \leq p_A < \tilde{p}_1$, Tit for Tat when $\tilde{p}_1 < p_A < \tilde{p}_2$, and Always Hawk when $\tilde{p}_2 < p_A \leq 1$. These critical values of $p_A$ are important for all of the different learning strategies; we give them for each parameter regime in Table \ref{T:CriticalP}.

In a weighted learning strategy, we construct the substrategy vector $\substratDoveProbs$ by taking a weighted mean of substrategy vectors for the different substrategies of PD TfT. The weighting for Always Dove is given by the probability that $p_A \in (0,\tilde{p}_1)$, the weighting for Tit for Tat is given by the probability that $p_A \in (\tilde{p}_1,\tilde{p}_2)$, and the weighting for Always Hawk is given by the probability that $p_A \in (\tilde{p}_2,1)$. 

In a mean-based learning strategy, $\substratDoveProbs$ is the substrategy of PD TfT that is associated with the mean of the distribution of an animal's current beliefs about $p_A$. The mean of $\mathrm{Beta}(\alpha,\beta)$ is $\frac{\alpha}{\alpha+ \beta}$, so animal A's strategy is
\begin{equation}
	\substratDoveProbs = 
\begin{cases}
\substratDoveProbs_\text{D}\,, & 0 \leq \frac{\alpha}{\alpha+\beta} < \tilde{p}_1\,, \\
\substratDoveProbs_\text{TfT}\,, & \tilde{p}_1 \leq \frac{\alpha}{\alpha+\beta} < \tilde{p}_2\,, \\
\substratDoveProbs_\text{H}\,, & \tilde{p}_2 \leq \frac{\alpha}{\alpha+\beta} \leq 1  \,,
\end{cases}
\end{equation}
where $\substratDoveProbs_\text{D}$,  $\substratDoveProbs_\text{TfT}$, and $\substratDoveProbs_\text{H}$ denote the substrategy vectors of `Always Dove', `Tit for Tat', and `Always Hawk', respectively.
To maximize the amount of information that an animal learns, we make the most aggressive available choice for $\substratDoveProbs$ when
$\frac{\alpha}{\alpha+\beta}$ is at a critical value. For example, when $\frac{\alpha}{\alpha+\beta} = \tilde{p}_2$, we use the Always Hawk substrategy, rather than Tit for Tat or Always Dove.

A median-based learning strategy is identical to a mean-based learning strategy, except that we use the median of the distribution $\mathrm{Beta}(\alpha,\beta)$ instead of the mean. High-quantile learning strategies also use the same principle, but they are based on some other quantile of the distribution of an animal's beliefs about $p_A$. By basing such a learning strategy on a high quantile, an animal can behave aggressively until it has enough information to have a prescribed amount of confidence that aggressive behaviour is suboptimal. For example, if animal $A$ pursues a learning strategy that is based on the $0.95$ quantile, it uses an Always Hawk substrategy until the $0.95$ quantile of $\mathrm{Beta}(\alpha,\beta)$ is below the critical value $\tilde{p}_2$. Therefore, the animal plays Hawk during every move until it is 95\% confident that $p_A < \tilde{p}_2$.

In addition to specifying the relationship between current beliefs and current substrategy, it is important to specify each animal's initial beliefs about $p_A$ (as represented by the parameters $\alpha_0$ and $\beta_0$). Because we assume that animals initially have no information about $p_A$, apart from the fact that $p_A \sim \mathrm{Beta}(k,k)$, it may appear natural to choose $\alpha_0 = \beta_0 = k$.
However, there are potential advantages for an animal to use a prior that reflects its ``optimism'' about its chances of winning fights (as reflected by a large value of $\alpha_0$ and/or a small value of $\beta_0$).
Such an animal engages in more fights in the early rounds of a contest to gain information about its RHP relative to that of its opponent. One possible result is weak animals sustaining avoidable costs that they would not obtain if using an unbiased prior, but these costs can be outweighed by the benefits to stronger animals that might otherwise have ``given up'' following an unlucky loss in an early fight. Consequently, the expected payoff (averaged over all values of $p_A$) may be higher for an animal that uses a biased prior with $\alpha_0 > \beta_0$, instead of an unbiased prior.

For the mean-based, median-based, and high-quantile learning strategies, we introduce two additional features to deal with problems that can arise when changing substrategies. The first feature ensures that sharing develops if both animals adopt a TfT substrategy after a period of being more aggressive. The second feature introduces some hysteresis that prevents frequent changes between the Always Hawk and TfT substrategies. We describe these features in detail in Appendix \ref{S:TfTTransitions}.  


\subsection{Tournament for comparing learning strategies}
\label{S:LearningTournament}

As in Section \ref{S:InformedTournament}, we use tournaments to compare the performance of several learning strategies against each other and against the $9$ simple memory-$1$ strategies from Table \ref{T:Memory1Strategies}.
Our results are tabulated in Appendix \ref{S:ResultsInformed}. We consider $16$ different learning strategies. For each type of learning strategy (weighted, mean-based, median-based, and high-quantile), we consider $4$ different strategies, which entail different levels of aggression.
A more aggressive animal is willing to pay a higher cost for information about RHPs, and it requires more evidence (in the form of defeats in fights) before it adopts a less aggressive substrategy. In Table \ref{T:LearningStratsTable}, we outline the $25$ learning strategies in our tournaments. We take $k=1$, so we draw $p_A$ from a uniform distribution. As with our tournaments in Section \ref{S:InformedTournament}, we 
obtain the same qualitative results when we instead use $k = {1}/{2}$ or $k = 2$.

For weighted, mean-based, and median-based learning strategies, we encode aggressiveness through the values of the shape parameters $(\alpha_0,\beta_0)$ that determine an animal's initial beliefs about its chances of winning a fight. We interpret these parameters biologically by comparing them with the true distribution $p_A \sim \mathrm{Beta}(k,k)$ for the probability of winning a fight. When the distribution $\mathrm{Beta}(\alpha_0,\beta_0)$ determines an animal's prior beliefs about winning a fight, the animal acts as if it believes that it has already won $\alpha_0 - k$ fights and lost $\beta_0 - k$ fights at the beginning of a contest. In our learning strategies, we fix $\beta_0 = k$ and modify the level of aggression by using different values of $\alpha_0$. Specifically, we consider $\alpha_0 = k$, $\alpha_0 = k+4$, $\alpha_0 = k+8$, and $\alpha_0 = k+12$.

As we described in Section \ref{S:LearningDefns}, mean-based and median-based learning strategies can include a secondary prior $\mathrm{Beta}(\bar{\alpha}_0,\bar{\beta}_0)$ to incorporate hysteresis into the process of changing substrategies.
We take $\bar{\alpha}_0 = \bar{\beta}_0 = k$ in all mean-based and median-based strategies. For $k=1$, this choice entails a flat prior when making decisions about increasing the level of aggression.

We use the same prior for all high-quantile strategies, for which we examine different aggression levels by using different quantiles of the distribution to reflect an animal's beliefs about its chances of winning a fight. For the four different aggression levels, we use the $0.8$, $0.9$, $0.95$, and $0.98$ quantiles. In the most aggressive strategy, which is based on the $0.98$ quantile, an animal uses an Always Hawk substrategy until it has a 2\% or lower confidence that $p_A$ is in the range where Always Hawk is the best substrategy. We use a secondary quantile that is lower than the main quantile (see Appendix \ref{S:TfTTransitions}) for deciding when to change from a less aggressive substrategy to a more aggressive one. We use $0.7$, $0.8$, $0.9$, and $0.95$ as the secondary quantiles for $0.8$, $0.9$, $0.95$, and $0.98$, respectively.

In Tables \ref{T:RegimeILearningColourOnly}, \ref{T:RegimeIILearningColourOnly}, and \ref{T:RegimeIIILearningColourOnly}, we present our results for our tournaments in parameter regimes I, II, and III, respectively. We indicate strategies using our notation from Table \ref{T:LearningStratsTable}. We colour the squares in Tables \ref{T:RegimeILearningColourOnly}--\ref{T:RegimeIIILearningColourOnly} according to the values of $Q(T,S) = [E(T,S) - E(S,S)]/[E(\text{AllH},\text{AllD})]$. See the supplementary material for the values of the mean payoffs $E(T,S)$. (They are also available at \url{https://bitbucket.org/CameronLHall/dominancesharingassessmentmatlab/}.) Each contest consists of $1000$ rounds, and we use the parameter values $k = 1$, $\gamma = 0.995$, and $c_W = 0.1$. To examine results in the three main parameter regimes, we consider $c_L = 0.2$, $c_L = 0.6$, and $c_L = 1.2$ in the three tournaments. In all tournaments, we use the variance-reduction techniques from Appendix \ref{S:VarianceReduction}.

Unlike in our tournaments in Section \ref{S:InformedTournament}, we do not observe any strong ESSs. This is not surprising, because learning strategies with similar parameters (or ones with similar structures, like the mean-based and median-based strategies) are likely to perform similarly, and differences may be smaller than the `noise' from simulations. Consequently, it is unlikely that a single learning strategy appears to be an unambiguous ESS based on our simulations. Additionally, it is difficult to apply the method of \citet{Haigh1975} to find mixed ESSs for this tournament, both because the game that is defined by the mean payoffs is nongeneric and because the time that is required for Haigh's algorithm grows exponentially in the number of strategies. Instead, we draw qualitative conclusions about the relative performance of different learning strategies by examining Tables \ref{T:RegimeILearningColourOnly}, \ref{T:RegimeIILearningColourOnly}, and \ref{T:RegimeIIILearningColourOnly}.

For the most part, we observe that the learning strategies (J--Y) are resistant to invasion by the simple strategies (A--I). The least aggressive (i.e., with $\alpha_0 = \beta_0 = 1$) mean-based and median-based strategies are less successful than the other learning strategies.
Weighted strategies are more successful than the least aggressive mean-based and median-based strategies, but they are less successful than other learning strategies, especially in parameter regime III. Additionally, the most aggressive mean-based, median-based, and high-quantile strategies are less successful (in the sense of being vulnerable to invasion) than some other learning strategies in parameter regime III. 

We also observe many white and light-coloured cells (especially in parameter regime I) when a pair of learning strategies interact with each other.
Therefore, it appears that a variety of different learning strategies may be able to coexist successfully. In a population in which animals use results from fights to inform their long-term behaviour, there may not be a single optimal strategy for using such information. Mean-based, median-based, and high-quantile learning strategies all have the potential to be effective if the aggression parameters (i.e., $\alpha_0$ for the mean-based and median-based strategies and the quantile for the high-quantile strategies) are not too large or too small. Parameter regime I appears to have the broadest range of parameter values that correspond to effective learning strategies, possibly because the penalty for fighting is relatively small.

One prominent feature of the results in Tables \ref{T:RegimeILearningColourOnly}--\ref{T:RegimeIIILearningColourOnly} is that every cell in column $G$ is either dark blue or white. This illustrates that the simple `Grim' strategy is very successful, although it is not a strong ESS, as evidenced by the white off-diagonal cells. These white cells indicate that mutant animals who use some other strategies (e.g., Always Dove or Tit for Tat) obtain the same payoff as animals in a population who use the Grim strategy, enabling the mutants to invade the population because of genetic drift. A population of animals that play Grim all do very well, as they share their resources during each round. However, a mutant animal that pursues a learning strategy in a population of animals who play Grim is very unsuccessful. When a mutant plays Hawk to obtain information about its probability of winning fights, the response of a Grim is to punish the mutant by playing Hawk in every subsequent round. Even if the mutant has a relatively large RHP, this is likely to lead to an overall payoff that is lower than what it would obtain by playing Dove in every round.

The columns that correspond to strategies O, P, T, W, and X have only white and blue cells in all parameter regimes. This indicates that these strategies are also difficult to invade. Like Grim, it is only possible to invade them via genetic drift. Strategies O, P, T, W, and X share the property that animals that pursuing these strategies begin by playing Hawk in every round and then modify their behaviour if they are unsuccessful in fights. This is consistent with the idea that successful animals without prior knowledge of $p_A$ use costly assessment (which we model as fighting) to establish their strength relative to their opponent, and they then adopt a long-term behaviour that is consistent with their estimate of their own strength. From simulations of contests between animals that pursue these strategies (not shown), we find that the assessment process takes longer (involving more rounds of fighting) when $p_A$ is close to a critical threshold value. This indicates that assessment is more costly when animals are more evenly matched.


\section{Conclusions and Discussion}
\label{S:Discussion}

We have developed a model of animal conflict based on an iterated Hawk--Dove game to analyze how different factors affect whether the outcome of a contest between animals is resource sharing, overt aggression, or the formation of a dominance relationship. We have shown that the same model can explain the conditions under which very different outcomes are evolutionarily stable and that key factors that determine which outcomes occur are costs of fighting, degree of asymmetry between animals in a contest, and the ability to learn from experience.

Although the Iterated Prisoners Dilemma and modifications of it \citep[e.g., see][]{AxelrodEvolCooperation, Nowak2006} have been cited extensively as evidence for the evolution of cooperation and resource sharing, our findings demonstrate that sharing is stable only under rather limited conditions --- specifically, when the cost of fighting is low and the animals in a contest have similar RHPs. As the asymmetries between animals become larger, resource sharing becomes less stable. 
By contrast, dominance hierarchies are stable for a much wider range of conditions. The explanation appears to be that progressively larger differences in RHPs make it progressively more beneficial both for the stronger animal in a contest to fight (rather than share) and for the weaker animal to demur without even displaying or attempting to fight.  
Consequently, dominance hierarchies develop readily in the face of asymmetries and become progressively more stable for progressively larger asymmetries \citep{Dawkins2010}. 
This provides an explanation for why dominance hierarchies are widespread in so many species \citep{HuntingfordAnimalConflict,Bonabeau1999,Braddock1955,Beacham1987,Chase1982,Drummond2006,Guhl1968,ODonnell1998}. 

When animals do not know RHPs at the outset of a contest, we found that the most successful strategies involve a limited period of assessment followed by a longer period in which they avoid fights and a stable dominance hierarchy arises. This is consistent with observations \citep{Chase1982,Pagel1997}. 
We also found that the duration of assessment depends both on the costliness of fighting and on the difference between the RHPs of the animals in a contest. Although animals that used assessment tended to outperform animals that used simple strategies, we also observed that Grim is resistant to invasion from mutants who use learning strategies. An interesting open problem is to determine whether 
it is possible for a mixture of simple strategies to be an ESS when some animals are capable of using assessment.

We have explored only a small number of the myriad strategies that are possible in our iterated asymmetric Hawk--Dove game. 
To promote future work, on we have developed software (see \url{https://bitbucket.org/CameronLHall/dominancesharingassessmentmatlab/}) to allow readers to pursue our ideas further and extend them to a wider range of questions. For example, our informed strategies from Section \ref{S:Informed} are examples of the memory-$1$ Markovian strategies in \citet{Nowak1993} and \citet{Press2012}.
An interesting extension is to consider memory-$1$ Markovian strategies for iterated asymmetric games more generally, following a similar approach to those in \citet{Nowak1990} and \citet{Nowak1993}. Other valuable extensions include the consideration of memory-$s$ strategies for $s \geq 2$, analysis of the impact of bifurcations in the dynamics of tournaments with noise, and the study of tournaments of animals on networks \citep{szabo2007}. 
We hope that the simplicity of our model and the ease of use of our software will enable the development of answers to these and many other questions about animal conflict.


\section*{Acknowledgements}

We thank Jasvir Grewal for early work on this project. CLH acknowledges support from the Mathematics Applications Consortium for Science and Industry (\href{https://ulsites.ul.ie/macsi}{https://ulsites.ul.ie/macsi}) funded by the Science Foundation Ireland grant investigator award 12/IA/1683.



\section*{References}



\section*{Appendices}

\appendix
\numberwithin{equation}{section}
\numberwithin{figure}{section}
\numberwithin{table}{section}


\section{Computation of tournament results}
\label{S:VarianceReduction}

To maximize the accuracy of our approximations of expected payoffs and ensure that it is reasonable to compare them between strategy pairs, we employ variance-reduction techniques that are common in Monte Carlo methods \citep{OwenMonteCarlo}. We begin each tournament by generating a set of values for $p_A$, and we use the same set of values of $p_A$ for each strategy-pair comparison. This is equivalent to using the method of common numbers for variance reduction \citep{OwenMonteCarlo}; it increases the likelihood that the differences between mean payoffs for different strategy pairs result from differences in the expected payoffs, rather than from random fluctuations. 
We also use the method of antithetic sampling \citep{OwenMonteCarlo} to minimize any bias in favor of animals $A$ or $B$ in the calculated mean payoffs.
We implement the method of antithetic sampling by ensuring, for each value of $p_A$, that we also use its complement $1 - p_A$ as a value for $p_A$.

In the tournaments in our paper, we draw $250$ values for $p_A$ from the uniform distribution $U(0,1) = \mathrm{Beta}(1,1)$, and we then generate an additional $250$ values of $p_A$ by taking the complements of the first set of values. For each strategy-pair comparison, we run $20$ contests with each of the $500$ values of $p_A$. This gives a total of $10,000$ contests for each strategy comparison; we report the mean payoff to animal $A$ from these $10,000$ contests. The purpose of performing multiple contests with each value of $p_A$ is to reduce the noise in our plots of mean payoff as a function of $p_A$. (We do not show these plots in this paper, but one can generate them using the code at \url{https://bitbucket.org/CameronLHall/dominancesharingassessmentmatlab/}.)


\section[Tournament results for animals with knowledge of pA]{Tournament results for animals with knowledge of $p_A$}
\label{S:ResultsInformed}

In Tables \ref{T:RegimeIGridOfOutputs}, \ref{T:RegimeIIGridOfOutputs}, and \ref{T:RegimeIIIGridOfOutputs}, we describe our results for simulations in parameter regimes I, II, and III, respectively. In these tables, we show the mean payoff to animal $A$ when it pursues the strategy in the row and animal $B$ pursues the strategy in the column. 

The colours indicate whether a mutant with a different strategy can successfully invade a population of animals who are using a single strategy. In these tables, a mutant pursues a strategy in the row and the population strategy is in the column. The colour of a cell indicates whether the expected payoff to the mutant is more than, the same as, or less than the expected payoff for an animal using the population strategy against another animal using the population strategy. A red cell indicates that a mutant has a higher payoff than animals in the population, so the row strategy can successfully invade the column strategy. By contrast, a blue cell indicates that the mutant has a lower payoff than animals in the population, so the column strategy is resistant to invasion by that row strategy. A white cell indicates that the mutant has a similar or identical payoff to the population, so genetic drift may lead to a mixed population with animals who pursue both strategies.

The intensity of the colours indicates the size of the difference between the expected payoff to a mutant and the expected payoff to an animal that uses the population strategy. Specifically, we colour cells according to $Q(T,S) = [E(T,S) - E(S,S)]/[E(\text{AllH},\text{AllD})]$, where $T$ is the strategy that is pursued by animal $A$ and $S$ is the strategy that is pursued by animal $B$. Cells are dark when $|Q(T,S)| > 10^{-1}$, of a medium shade when $10^{-2} <  |Q(T,S)| \leq 10^{-1}$, and light when $10^{-3} < |Q(T,S)| < 10^{-2}$.

\begin{landscape}
	\begin{table}[ht]
		\centering
		\footnotesize 
\setlength\tabcolsep{4pt}
\begin{tabular}{|l|llllllllllll|}
\hline
      & A     & B     & C     & D     & E     & F     & G     & H     & I     & J     & K     & L \bigstrut\\
\hline
(A) AllH & 69.5\phantom{1}  & \cellcolor[rgb]{ .937,  .541,  .384}101.8 & \cellcolor[rgb]{ .698,  .094,  .169}\textcolor[rgb]{ 1,  1,  1}{134.1} & \cellcolor[rgb]{ .698,  .094,  .169}\textcolor[rgb]{ 1,  1,  1}{166.4} & \cellcolor[rgb]{ .698,  .094,  .169}\textcolor[rgb]{ 1,  1,  1}{198.7} & \cellcolor[rgb]{ .129,  .4,  .675}\textcolor[rgb]{ 1,  1,  1}{70.1} & \cellcolor[rgb]{ .129,  .4,  .675}\textcolor[rgb]{ 1,  1,  1}{70.2} & \cellcolor[rgb]{ .698,  .094,  .169}\textcolor[rgb]{ 1,  1,  1}{134.2} & \cellcolor[rgb]{ .937,  .541,  .384}102.2 & \cellcolor[rgb]{ .937,  .541,  .384}108.3 & \cellcolor[rgb]{ .404,  .663,  .812}75.5\phantom{1} & \cellcolor[rgb]{ .404,  .663,  .812}75.8\phantom{1} \bigstrut[t]\\
(B) 25\%D & \cellcolor[rgb]{ .404,  .663,  .812}52.2 & 82.6  & \cellcolor[rgb]{ .698,  .094,  .169}\textcolor[rgb]{ 1,  1,  1}{113.0} & \cellcolor[rgb]{ .698,  .094,  .169}\textcolor[rgb]{ 1,  1,  1}{143.4} & \cellcolor[rgb]{ .698,  .094,  .169}\textcolor[rgb]{ 1,  1,  1}{173.8} & \cellcolor[rgb]{ .404,  .663,  .812}83.0 & \cellcolor[rgb]{ .129,  .4,  .675}\textcolor[rgb]{ 1,  1,  1}{53.0} & \cellcolor[rgb]{ .937,  .541,  .384}113.0 & \cellcolor[rgb]{ .404,  .663,  .812}84.6 & \cellcolor[rgb]{ .404,  .663,  .812}93.7 & \cellcolor[rgb]{ .404,  .663,  .812}60.5 & \cellcolor[rgb]{ .404,  .663,  .812}73.7 \\
(C) 50\%D & \cellcolor[rgb]{ .129,  .4,  .675}\textcolor[rgb]{ 1,  1,  1}{34.8} & \cellcolor[rgb]{ .404,  .663,  .812}63.3 & 91.9  & \cellcolor[rgb]{ .698,  .094,  .169}\textcolor[rgb]{ 1,  1,  1}{120.4} & \cellcolor[rgb]{ .698,  .094,  .169}\textcolor[rgb]{ 1,  1,  1}{149.0} & \cellcolor[rgb]{ .404,  .663,  .812}92.2 & \cellcolor[rgb]{ .129,  .4,  .675}\textcolor[rgb]{ 1,  1,  1}{35.9} & \cellcolor[rgb]{ .404,  .663,  .812}91.9 & \cellcolor[rgb]{ .129,  .4,  .675}\textcolor[rgb]{ 1,  1,  1}{67.3} & \cellcolor[rgb]{ .129,  .4,  .675}\textcolor[rgb]{ 1,  1,  1}{79.0} & \cellcolor[rgb]{ .129,  .4,  .675}\textcolor[rgb]{ 1,  1,  1}{45.5} & \cellcolor[rgb]{ .404,  .663,  .812}71.2 \\
(D) 75\%D & \cellcolor[rgb]{ .129,  .4,  .675}\textcolor[rgb]{ 1,  1,  1}{17.4} & \cellcolor[rgb]{ .129,  .4,  .675}\textcolor[rgb]{ 1,  1,  1}{44.1} & \cellcolor[rgb]{ .129,  .4,  .675}\textcolor[rgb]{ 1,  1,  1}{70.8} & 97.5  & \cellcolor[rgb]{ .698,  .094,  .169}\textcolor[rgb]{ 1,  1,  1}{124.2} & \cellcolor[rgb]{ .82,  .898,  .941}97.6 & \cellcolor[rgb]{ .129,  .4,  .675}\textcolor[rgb]{ 1,  1,  1}{19.5} & \cellcolor[rgb]{ .129,  .4,  .675}\textcolor[rgb]{ 1,  1,  1}{70.8} & \cellcolor[rgb]{ .129,  .4,  .675}\textcolor[rgb]{ 1,  1,  1}{50.2} & \cellcolor[rgb]{ .129,  .4,  .675}\textcolor[rgb]{ 1,  1,  1}{64.3} & \cellcolor[rgb]{ .129,  .4,  .675}\textcolor[rgb]{ 1,  1,  1}{30.4} & \cellcolor[rgb]{ .404,  .663,  .812}67.9 \\
(E) AllH & \cellcolor[rgb]{ .129,  .4,  .675}\textcolor[rgb]{ 1,  1,  1}{0.0} & \cellcolor[rgb]{ .129,  .4,  .675}\textcolor[rgb]{ 1,  1,  1}{24.8} & \cellcolor[rgb]{ .129,  .4,  .675}\textcolor[rgb]{ 1,  1,  1}{49.7} & \cellcolor[rgb]{ .129,  .4,  .675}\textcolor[rgb]{ 1,  1,  1}{74.5} & 99.3  & 99.3  & 99.3  & \cellcolor[rgb]{ .129,  .4,  .675}\textcolor[rgb]{ 1,  1,  1}{49.8} & \cellcolor[rgb]{ .129,  .4,  .675}\textcolor[rgb]{ 1,  1,  1}{49.6} & \cellcolor[rgb]{ .129,  .4,  .675}\textcolor[rgb]{ 1,  1,  1}{49.7} & \cellcolor[rgb]{ .129,  .4,  .675}\textcolor[rgb]{ 1,  1,  1}{15.3} & \cellcolor[rgb]{ .129,  .4,  .675}\textcolor[rgb]{ 1,  1,  1}{64.2} \\
(F) TfT & \cellcolor[rgb]{ .82,  .898,  .941}69.2 & \cellcolor[rgb]{ .82,  .898,  .941}82.3 & \cellcolor[rgb]{ .82,  .898,  .941}91.6 & 97.4  & 99.3  & 99.3  & 99.3  & \cellcolor[rgb]{ .404,  .663,  .812}92.3 & \cellcolor[rgb]{ .404,  .663,  .812}79.2 & \cellcolor[rgb]{ .129,  .4,  .675}\textcolor[rgb]{ 1,  1,  1}{58.7} & \cellcolor[rgb]{ .404,  .663,  .812}60.1 & \cellcolor[rgb]{ .129,  .4,  .675}\textcolor[rgb]{ 1,  1,  1}{65.0} \\
(G) Grim & \cellcolor[rgb]{ .82,  .898,  .941}69.3 & \cellcolor[rgb]{ .937,  .541,  .384}101.3 & \cellcolor[rgb]{ .698,  .094,  .169}\textcolor[rgb]{ 1,  1,  1}{133.2} & \cellcolor[rgb]{ .698,  .094,  .169}\textcolor[rgb]{ 1,  1,  1}{164.6} & 99.3  & 99.3  & 99.3  & \cellcolor[rgb]{ .698,  .094,  .169}\textcolor[rgb]{ 1,  1,  1}{127.6} & \cellcolor[rgb]{ .937,  .541,  .384}101.2 & \cellcolor[rgb]{ .129,  .4,  .675}\textcolor[rgb]{ 1,  1,  1}{58.7} & \cellcolor[rgb]{ .404,  .663,  .812}60.0 & \cellcolor[rgb]{ .129,  .4,  .675}\textcolor[rgb]{ 1,  1,  1}{65.0} \\
(H) Pav & \cellcolor[rgb]{ .129,  .4,  .675}\textcolor[rgb]{ 1,  1,  1}{34.7} & \cellcolor[rgb]{ .404,  .663,  .812}63.3 & 91.9  & \cellcolor[rgb]{ .698,  .094,  .169}\textcolor[rgb]{ 1,  1,  1}{120.4} & \cellcolor[rgb]{ .698,  .094,  .169}\textcolor[rgb]{ 1,  1,  1}{148.7} & \cellcolor[rgb]{ .404,  .663,  .812}92.8 & \cellcolor[rgb]{ .129,  .4,  .675}\textcolor[rgb]{ 1,  1,  1}{46.3} & 98.6  & \cellcolor[rgb]{ .937,  .541,  .384}90.2 & \cellcolor[rgb]{ .129,  .4,  .675}\textcolor[rgb]{ 1,  1,  1}{79.2} & \cellcolor[rgb]{ .129,  .4,  .675}\textcolor[rgb]{ 1,  1,  1}{45.5} & \cellcolor[rgb]{ .404,  .663,  .812}71.3 \\
(I) MPav & \cellcolor[rgb]{ .404,  .663,  .812}55.5 & \cellcolor[rgb]{ .992,  .859,  .78}83.3 & \cellcolor[rgb]{ .937,  .541,  .384}111.1 & \cellcolor[rgb]{ .698,  .094,  .169}\textcolor[rgb]{ 1,  1,  1}{138.5} & \cellcolor[rgb]{ .698,  .094,  .169}\textcolor[rgb]{ 1,  1,  1}{149.0} & \cellcolor[rgb]{ .404,  .663,  .812}96.0 & \cellcolor[rgb]{ .129,  .4,  .675}\textcolor[rgb]{ 1,  1,  1}{63.6} & \cellcolor[rgb]{ .937,  .541,  .384}104.6 & 87.5  & \cellcolor[rgb]{ .404,  .663,  .812}80.6 & \cellcolor[rgb]{ .129,  .4,  .675}\textcolor[rgb]{ 1,  1,  1}{56.0} & \cellcolor[rgb]{ .404,  .663,  .812}71.7 \\
(J) Bul & \cellcolor[rgb]{ .404,  .663,  .812}60.5 & 82.6  & \cellcolor[rgb]{ .937,  .541,  .384}104.8 & \cellcolor[rgb]{ .698,  .094,  .169}\textcolor[rgb]{ 1,  1,  1}{126.9} & \cellcolor[rgb]{ .698,  .094,  .169}\textcolor[rgb]{ 1,  1,  1}{149.0} & \cellcolor[rgb]{ .937,  .541,  .384}110.4 & \cellcolor[rgb]{ .937,  .541,  .384}110.4 & \cellcolor[rgb]{ .937,  .541,  .384}104.5 & \cellcolor[rgb]{ .937,  .541,  .384}101.0 & 99.3  & \cellcolor[rgb]{ .404,  .663,  .812}66.6 & \cellcolor[rgb]{ .404,  .663,  .812}81.3 \\
(K) N, MN, SDT & \cellcolor[rgb]{ .937,  .541,  .384}72.7 & \cellcolor[rgb]{ .937,  .541,  .384}100.3 & \cellcolor[rgb]{ .698,  .094,  .169}\textcolor[rgb]{ 1,  1,  1}{128.1} & \cellcolor[rgb]{ .698,  .094,  .169}\textcolor[rgb]{ 1,  1,  1}{155.7} & \cellcolor[rgb]{ .698,  .094,  .169}\textcolor[rgb]{ 1,  1,  1}{183.4} & \cellcolor[rgb]{ .404,  .663,  .812}88.5 & \cellcolor[rgb]{ .404,  .663,  .812}88.4 & \cellcolor[rgb]{ .698,  .094,  .169}\textcolor[rgb]{ 1,  1,  1}{127.8} & \cellcolor[rgb]{ .698,  .094,  .169}\textcolor[rgb]{ 1,  1,  1}{110.2} & \cellcolor[rgb]{ .937,  .541,  .384}111.5 & 78.7  & \cellcolor[rgb]{ .404,  .663,  .812}79.0 \\
(L) Slf, PDT, InT \, & \cellcolor[rgb]{ .937,  .541,  .384}72.5 & \cellcolor[rgb]{ .937,  .541,  .384}92.6 & \cellcolor[rgb]{ .937,  .541,  .384}109.7 & \cellcolor[rgb]{ .698,  .094,  .169}\textcolor[rgb]{ 1,  1,  1}{123.6} & \cellcolor[rgb]{ .698,  .094,  .169}\textcolor[rgb]{ 1,  1,  1}{134.5} & \cellcolor[rgb]{ .937,  .541,  .384}112.6 & \cellcolor[rgb]{ .937,  .541,  .384}112.6 & \cellcolor[rgb]{ .937,  .541,  .384}110.2 & \cellcolor[rgb]{ .937,  .541,  .384}103.2 & \cellcolor[rgb]{ .404,  .663,  .812}97.0 & 78.6  & 87.5 \bigstrut[b]\\
\hline
\end{tabular}%

		\normalsize
		\caption{Overall payoffs to animal $A$ in parameter regime I when it pursues the strategy in
			the row and animal $B$ pursues the strategy in the column.
			Our parameter values are $c_W = 0.1$, $c_L = 0.2$, $\gamma = 0.995$, and $k = 1$.
			We colour cells according to $Q(T,S) = [E(T,S) - E(S,S)]/[E(\text{AllH},\text{AllD})]$, where $T$ is the strategy that is pursued by animal $A$ (indicated by the row) and $S$ is the strategy that is pursued by animal $B$ (indicated by the column).
			Cells are red when $Q(T,S) > 10^{-3}$, blue when $Q(T,S) < -10^{-3}$, and white when $|Q(T,S)| \leq 10^{-3}$.
			Cells are dark when $|Q(T,S)| > 10^{-1}$, of a medium shade when $10^{-2} <  |Q(T,S)| \leq 10^{-1}$, and light when $10^{-3} < |Q(T,S)| < 10^{-2}$. The column strategies correspond to the row strategies with the same letter.
			We assign a common letter to strategies that are identical in parameter regime I.
			We abbreviate strategies as follows: AllH = Always Hawk; X\%D = X\% Dove and otherwise Hawk; AllD = Always Dove; TfT  = Tit for Tat; Pav = Pavlov; MPav = Modified Pavlov; Bul = Bully; N = Nash; PDT = PD Tit for Tat; MN = Mixed Nash; Slf = Selfish; SDT = Snowdrift Tit for Tat; and InT = Informed Tit for Tat.
			\label{T:RegimeIGridOfOutputs}
		}
	\end{table}

	\begin{table}[ht]
		\centering
		\footnotesize
\setlength\tabcolsep{3pt}
\begin{tabular}{|l|lllllllllllllll|}
\hline
      & A     & B     & C     & D     & E     & F     & G     & H     & I     & J     & K     & L     & M     & N     & O \bigstrut\\
\hline
(A) AllH & 29.7\phantom{1}  & \cellcolor[rgb]{ .937,  .541,  .384}72.0\phantom{1} & \cellcolor[rgb]{ .698,  .094,  .169}\textcolor[rgb]{ 1,  1,  1}{114.2} & \cellcolor[rgb]{ .698,  .094,  .169}\textcolor[rgb]{ 1,  1,  1}{156.4} & \cellcolor[rgb]{ .698,  .094,  .169}\textcolor[rgb]{ 1,  1,  1}{198.7} & \cellcolor[rgb]{ .129,  .4,  .675}\textcolor[rgb]{ 1,  1,  1}{30.6} & \cellcolor[rgb]{ .129,  .4,  .675}\textcolor[rgb]{ 1,  1,  1}{30.7} & \cellcolor[rgb]{ .937,  .541,  .384}114.2 & \cellcolor[rgb]{ .82,  .898,  .941}70.8\phantom{1} & \cellcolor[rgb]{ .129,  .4,  .675}\textcolor[rgb]{ 1,  1,  1}{77.4} & \cellcolor[rgb]{ .129,  .4,  .675}\textcolor[rgb]{ 1,  1,  1}{62.5}\phantom{1} & \cellcolor[rgb]{ .129,  .4,  .675}\textcolor[rgb]{ 1,  1,  1}{62.9} & \cellcolor[rgb]{ .129,  .4,  .675}\textcolor[rgb]{ 1,  1,  1}{43.2}\phantom{1} & \cellcolor[rgb]{ .129,  .4,  .675}\textcolor[rgb]{ 1,  1,  1}{62.7}\phantom{1} & \cellcolor[rgb]{ .129,  .4,  .675}\textcolor[rgb]{ 1,  1,  1}{43.7}\phantom{1} \bigstrut[t]\\
(B) 25\%D & \cellcolor[rgb]{ .404,  .663,  .812}22.4 & 60.2  & \cellcolor[rgb]{ .937,  .541,  .384}98.1 & \cellcolor[rgb]{ .698,  .094,  .169}\textcolor[rgb]{ 1,  1,  1}{136.0} & \cellcolor[rgb]{ .698,  .094,  .169}\textcolor[rgb]{ 1,  1,  1}{173.8} & \cellcolor[rgb]{ .129,  .4,  .675}\textcolor[rgb]{ 1,  1,  1}{60.7} & \cellcolor[rgb]{ .129,  .4,  .675}\textcolor[rgb]{ 1,  1,  1}{23.3} & \cellcolor[rgb]{ .992,  .859,  .78}98.1 & \cellcolor[rgb]{ .404,  .663,  .812}60.9 & \cellcolor[rgb]{ .129,  .4,  .675}\textcolor[rgb]{ 1,  1,  1}{70.5} & \cellcolor[rgb]{ .129,  .4,  .675}\textcolor[rgb]{ 1,  1,  1}{56.9} & \cellcolor[rgb]{ .129,  .4,  .675}\textcolor[rgb]{ 1,  1,  1}{71.9} & \cellcolor[rgb]{ .129,  .4,  .675}\textcolor[rgb]{ 1,  1,  1}{43.1} & \cellcolor[rgb]{ .129,  .4,  .675}\textcolor[rgb]{ 1,  1,  1}{64.3} & \cellcolor[rgb]{ .129,  .4,  .675}\textcolor[rgb]{ 1,  1,  1}{58.1} \\
(C) 50\%D & \cellcolor[rgb]{ .404,  .663,  .812}14.9 & \cellcolor[rgb]{ .404,  .663,  .812}48.4 & 81.9  & \cellcolor[rgb]{ .698,  .094,  .169}\textcolor[rgb]{ 1,  1,  1}{115.5} & \cellcolor[rgb]{ .698,  .094,  .169}\textcolor[rgb]{ 1,  1,  1}{149.0} & \cellcolor[rgb]{ .404,  .663,  .812}82.3 & \cellcolor[rgb]{ .129,  .4,  .675}\textcolor[rgb]{ 1,  1,  1}{16.3} & \cellcolor[rgb]{ .404,  .663,  .812}81.9 & \cellcolor[rgb]{ .129,  .4,  .675}\textcolor[rgb]{ 1,  1,  1}{51.3} & \cellcolor[rgb]{ .129,  .4,  .675}\textcolor[rgb]{ 1,  1,  1}{63.5} & \cellcolor[rgb]{ .129,  .4,  .675}\textcolor[rgb]{ 1,  1,  1}{51.3} & \cellcolor[rgb]{ .129,  .4,  .675}\textcolor[rgb]{ 1,  1,  1}{77.0} & \cellcolor[rgb]{ .129,  .4,  .675}\textcolor[rgb]{ 1,  1,  1}{42.6} & \cellcolor[rgb]{ .129,  .4,  .675}\textcolor[rgb]{ 1,  1,  1}{64.3} & \cellcolor[rgb]{ .129,  .4,  .675}\textcolor[rgb]{ 1,  1,  1}{68.2} \\
(D) 75\%D & \cellcolor[rgb]{ .129,  .4,  .675}\textcolor[rgb]{ 1,  1,  1}{7.5} & \cellcolor[rgb]{ .129,  .4,  .675}\textcolor[rgb]{ 1,  1,  1}{36.7} & \cellcolor[rgb]{ .404,  .663,  .812}65.7 & 95.0  & \cellcolor[rgb]{ .698,  .094,  .169}\textcolor[rgb]{ 1,  1,  1}{124.2} & \cellcolor[rgb]{ .404,  .663,  .812}95.2 & \cellcolor[rgb]{ .129,  .4,  .675}\textcolor[rgb]{ 1,  1,  1}{9.7} & \cellcolor[rgb]{ .129,  .4,  .675}\textcolor[rgb]{ 1,  1,  1}{66.0} & \cellcolor[rgb]{ .129,  .4,  .675}\textcolor[rgb]{ 1,  1,  1}{42.2} & \cellcolor[rgb]{ .129,  .4,  .675}\textcolor[rgb]{ 1,  1,  1}{56.6} & \cellcolor[rgb]{ .129,  .4,  .675}\textcolor[rgb]{ 1,  1,  1}{45.7} & \cellcolor[rgb]{ .129,  .4,  .675}\textcolor[rgb]{ 1,  1,  1}{78.1} & \cellcolor[rgb]{ .129,  .4,  .675}\textcolor[rgb]{ 1,  1,  1}{41.6} & \cellcolor[rgb]{ .129,  .4,  .675}\textcolor[rgb]{ 1,  1,  1}{62.6} & \cellcolor[rgb]{ .129,  .4,  .675}\textcolor[rgb]{ 1,  1,  1}{73.9} \\
(E) AllH & \cellcolor[rgb]{ .129,  .4,  .675}\textcolor[rgb]{ 1,  1,  1}{0.0} & \cellcolor[rgb]{ .129,  .4,  .675}\textcolor[rgb]{ 1,  1,  1}{24.8} & \cellcolor[rgb]{ .129,  .4,  .675}\textcolor[rgb]{ 1,  1,  1}{49.7} & \cellcolor[rgb]{ .129,  .4,  .675}\textcolor[rgb]{ 1,  1,  1}{74.5} & 99.3  & 99.3  & 99.3  & \cellcolor[rgb]{ .129,  .4,  .675}\textcolor[rgb]{ 1,  1,  1}{49.5} & \cellcolor[rgb]{ .129,  .4,  .675}\textcolor[rgb]{ 1,  1,  1}{49.7} & \cellcolor[rgb]{ .129,  .4,  .675}\textcolor[rgb]{ 1,  1,  1}{49.7} & \cellcolor[rgb]{ .129,  .4,  .675}\textcolor[rgb]{ 1,  1,  1}{40.1} & \cellcolor[rgb]{ .129,  .4,  .675}\textcolor[rgb]{ 1,  1,  1}{75.1} & \cellcolor[rgb]{ .129,  .4,  .675}\textcolor[rgb]{ 1,  1,  1}{40.1} & \cellcolor[rgb]{ .129,  .4,  .675}\textcolor[rgb]{ 1,  1,  1}{59.2} & \cellcolor[rgb]{ .129,  .4,  .675}\textcolor[rgb]{ 1,  1,  1}{75.1} \\
(F) TfT & 29.7  & 60.1  & 81.8  & 94.9  & 99.3  & 99.3  & 99.3  & \cellcolor[rgb]{ .404,  .663,  .812}83.3 & \cellcolor[rgb]{ .404,  .663,  .812}58.8 & \cellcolor[rgb]{ .129,  .4,  .675}\textcolor[rgb]{ 1,  1,  1}{27.8} & \cellcolor[rgb]{ .129,  .4,  .675}\textcolor[rgb]{ 1,  1,  1}{22.5} & \cellcolor[rgb]{ .129,  .4,  .675}\textcolor[rgb]{ 1,  1,  1}{54.7} & \cellcolor[rgb]{ .129,  .4,  .675}\textcolor[rgb]{ 1,  1,  1}{22.6} & \cellcolor[rgb]{ .129,  .4,  .675}\textcolor[rgb]{ 1,  1,  1}{35.9} & \cellcolor[rgb]{ .129,  .4,  .675}\textcolor[rgb]{ 1,  1,  1}{54.7} \\
(G) Grim & 29.6  & \cellcolor[rgb]{ .937,  .541,  .384}71.7 & \cellcolor[rgb]{ .698,  .094,  .169}\textcolor[rgb]{ 1,  1,  1}{113.7} & \cellcolor[rgb]{ .698,  .094,  .169}\textcolor[rgb]{ 1,  1,  1}{154.8} & 99.3  & 99.3  & 99.3  & \cellcolor[rgb]{ .937,  .541,  .384}111.1 & \cellcolor[rgb]{ .937,  .541,  .384}75.5 & \cellcolor[rgb]{ .129,  .4,  .675}\textcolor[rgb]{ 1,  1,  1}{27.9} & \cellcolor[rgb]{ .129,  .4,  .675}\textcolor[rgb]{ 1,  1,  1}{22.5} & \cellcolor[rgb]{ .129,  .4,  .675}\textcolor[rgb]{ 1,  1,  1}{54.7} & \cellcolor[rgb]{ .129,  .4,  .675}\textcolor[rgb]{ 1,  1,  1}{22.5} & \cellcolor[rgb]{ .129,  .4,  .675}\textcolor[rgb]{ 1,  1,  1}{35.9} & \cellcolor[rgb]{ .129,  .4,  .675}\textcolor[rgb]{ 1,  1,  1}{54.7} \\
(H) Pav & \cellcolor[rgb]{ .404,  .663,  .812}14.9 & \cellcolor[rgb]{ .404,  .663,  .812}48.4 & 82.0  & \cellcolor[rgb]{ .698,  .094,  .169}\textcolor[rgb]{ 1,  1,  1}{115.5} & \cellcolor[rgb]{ .698,  .094,  .169}\textcolor[rgb]{ 1,  1,  1}{148.8} & \cellcolor[rgb]{ .404,  .663,  .812}83.5 & \cellcolor[rgb]{ .129,  .4,  .675}\textcolor[rgb]{ 1,  1,  1}{29.7} & 97.7  & \cellcolor[rgb]{ .937,  .541,  .384}84.9 & \cellcolor[rgb]{ .129,  .4,  .675}\textcolor[rgb]{ 1,  1,  1}{63.2} & \cellcolor[rgb]{ .129,  .4,  .675}\textcolor[rgb]{ 1,  1,  1}{51.1} & \cellcolor[rgb]{ .129,  .4,  .675}\textcolor[rgb]{ 1,  1,  1}{77.7} & \cellcolor[rgb]{ .129,  .4,  .675}\textcolor[rgb]{ 1,  1,  1}{42.7} & \cellcolor[rgb]{ .129,  .4,  .675}\textcolor[rgb]{ 1,  1,  1}{64.4} & \cellcolor[rgb]{ .129,  .4,  .675}\textcolor[rgb]{ 1,  1,  1}{68.9} \\
(I) MPav & \cellcolor[rgb]{ .937,  .541,  .384}32.0 & \cellcolor[rgb]{ .937,  .541,  .384}66.0 & \cellcolor[rgb]{ .937,  .541,  .384}99.5 & \cellcolor[rgb]{ .698,  .094,  .169}\textcolor[rgb]{ 1,  1,  1}{132.9} & \cellcolor[rgb]{ .698,  .094,  .169}\textcolor[rgb]{ 1,  1,  1}{148.9} & \cellcolor[rgb]{ .404,  .663,  .812}84.0 & \cellcolor[rgb]{ .129,  .4,  .675}\textcolor[rgb]{ 1,  1,  1}{44.3} & \cellcolor[rgb]{ .937,  .541,  .384}103.7 & 72.0  & \cellcolor[rgb]{ .129,  .4,  .675}\textcolor[rgb]{ 1,  1,  1}{62.8} & \cellcolor[rgb]{ .129,  .4,  .675}\textcolor[rgb]{ 1,  1,  1}{51.7} & \cellcolor[rgb]{ .129,  .4,  .675}\textcolor[rgb]{ 1,  1,  1}{74.3} & \cellcolor[rgb]{ .129,  .4,  .675}\textcolor[rgb]{ 1,  1,  1}{42.2} & \cellcolor[rgb]{ .129,  .4,  .675}\textcolor[rgb]{ 1,  1,  1}{61.9} & \cellcolor[rgb]{ .129,  .4,  .675}\textcolor[rgb]{ 1,  1,  1}{64.6} \\
(J) Bul & \cellcolor[rgb]{ .698,  .094,  .169}\textcolor[rgb]{ 1,  1,  1}{51.7} & \cellcolor[rgb]{ .937,  .541,  .384}76.0 & \cellcolor[rgb]{ .937,  .541,  .384}100.3 & \cellcolor[rgb]{ .698,  .094,  .169}\textcolor[rgb]{ 1,  1,  1}{124.6} & \cellcolor[rgb]{ .698,  .094,  .169}\textcolor[rgb]{ 1,  1,  1}{149.0} & \cellcolor[rgb]{ .937,  .541,  .384}101.6 & \cellcolor[rgb]{ .937,  .541,  .384}101.6 & \cellcolor[rgb]{ .937,  .541,  .384}100.2 & \cellcolor[rgb]{ .698,  .094,  .169}\textcolor[rgb]{ 1,  1,  1}{95.7} & 99.3  & \cellcolor[rgb]{ .82,  .898,  .941}84.5 & \cellcolor[rgb]{ .937,  .541,  .384}110.0 & 65.2  & \cellcolor[rgb]{ .404,  .663,  .812}94.1 & \cellcolor[rgb]{ .404,  .663,  .812}90.7 \\
(K) N, MN\, & \cellcolor[rgb]{ .698,  .094,  .169}\textcolor[rgb]{ 1,  1,  1}{53.2} & \cellcolor[rgb]{ .937,  .541,  .384}79.6 & \cellcolor[rgb]{ .698,  .094,  .169}\textcolor[rgb]{ 1,  1,  1}{105.9} & \cellcolor[rgb]{ .698,  .094,  .169}\textcolor[rgb]{ 1,  1,  1}{132.2} & \cellcolor[rgb]{ .698,  .094,  .169}\textcolor[rgb]{ 1,  1,  1}{158.5} & \cellcolor[rgb]{ .404,  .663,  .812}93.8 & \cellcolor[rgb]{ .404,  .663,  .812}93.7 & \cellcolor[rgb]{ .937,  .541,  .384}106.0 & \cellcolor[rgb]{ .698,  .094,  .169}\textcolor[rgb]{ 1,  1,  1}{97.8} & \cellcolor[rgb]{ .992,  .859,  .78}100.8 & 86.0  & \cellcolor[rgb]{ .937,  .541,  .384}102.0 & \cellcolor[rgb]{ .937,  .541,  .384}66.7 & \cellcolor[rgb]{ .404,  .663,  .812}86.1 & \cellcolor[rgb]{ .404,  .663,  .812}82.8 \\
(L) Slf & \cellcolor[rgb]{ .698,  .094,  .169}\textcolor[rgb]{ 1,  1,  1}{53.1} & \cellcolor[rgb]{ .937,  .541,  .384}73.8 & \cellcolor[rgb]{ .937,  .541,  .384}92.5 & \cellcolor[rgb]{ .937,  .541,  .384}109.1 & \cellcolor[rgb]{ .698,  .094,  .169}\textcolor[rgb]{ 1,  1,  1}{123.6} & \cellcolor[rgb]{ .937,  .541,  .384}110.2 & \cellcolor[rgb]{ .937,  .541,  .384}110.2 & \cellcolor[rgb]{ .404,  .663,  .812}93.0 & \cellcolor[rgb]{ .937,  .541,  .384}90.5 & \cellcolor[rgb]{ .129,  .4,  .675}\textcolor[rgb]{ 1,  1,  1}{75.4} & \cellcolor[rgb]{ .404,  .663,  .812}70.0 & 99.3  & \cellcolor[rgb]{ .937,  .541,  .384}70.1 & \cellcolor[rgb]{ .404,  .663,  .812}83.4 & 99.3 \\
(M) SDT & \cellcolor[rgb]{ .698,  .094,  .169}\textcolor[rgb]{ 1,  1,  1}{50.4} & \cellcolor[rgb]{ .698,  .094,  .169}\textcolor[rgb]{ 1,  1,  1}{81.0} & \cellcolor[rgb]{ .698,  .094,  .169}\textcolor[rgb]{ 1,  1,  1}{109.1} & \cellcolor[rgb]{ .698,  .094,  .169}\textcolor[rgb]{ 1,  1,  1}{135.0} & \cellcolor[rgb]{ .698,  .094,  .169}\textcolor[rgb]{ 1,  1,  1}{158.5} & \cellcolor[rgb]{ .404,  .663,  .812}93.7 & \cellcolor[rgb]{ .404,  .663,  .812}93.6 & \cellcolor[rgb]{ .937,  .541,  .384}109.3 & \cellcolor[rgb]{ .698,  .094,  .169}\textcolor[rgb]{ 1,  1,  1}{97.6} & \cellcolor[rgb]{ .82,  .898,  .941}98.0 & \cellcolor[rgb]{ .404,  .663,  .812}83.1 & \cellcolor[rgb]{ .937,  .541,  .384}102.0 & 63.9  & \cellcolor[rgb]{ .404,  .663,  .812}83.3 & \cellcolor[rgb]{ .404,  .663,  .812}82.7 \\
(N) PDT & \cellcolor[rgb]{ .698,  .094,  .169}\textcolor[rgb]{ 1,  1,  1}{53.1} & \cellcolor[rgb]{ .937,  .541,  .384}77.3 & \cellcolor[rgb]{ .937,  .541,  .384}99.7 & \cellcolor[rgb]{ .698,  .094,  .169}\textcolor[rgb]{ 1,  1,  1}{120.4} & \cellcolor[rgb]{ .698,  .094,  .169}\textcolor[rgb]{ 1,  1,  1}{139.5} & \cellcolor[rgb]{ .937,  .541,  .384}106.9 & \cellcolor[rgb]{ .937,  .541,  .384}106.9 & \cellcolor[rgb]{ .937,  .541,  .384}99.9 & \cellcolor[rgb]{ .698,  .094,  .169}\textcolor[rgb]{ 1,  1,  1}{95.6} & \cellcolor[rgb]{ .404,  .663,  .812}91.3 & 85.9  & \cellcolor[rgb]{ .937,  .541,  .384}115.2 & \cellcolor[rgb]{ .937,  .541,  .384}66.7 & 99.3  & \cellcolor[rgb]{ .404,  .663,  .812}96.0 \\
(O) InT & \cellcolor[rgb]{ .698,  .094,  .169}\textcolor[rgb]{ 1,  1,  1}{50.2} & \cellcolor[rgb]{ .937,  .541,  .384}75.3 & \cellcolor[rgb]{ .937,  .541,  .384}95.8 & \cellcolor[rgb]{ .937,  .541,  .384}111.9 & \cellcolor[rgb]{ .698,  .094,  .169}\textcolor[rgb]{ 1,  1,  1}{123.6} & \cellcolor[rgb]{ .937,  .541,  .384}110.2 & \cellcolor[rgb]{ .937,  .541,  .384}110.2 & \cellcolor[rgb]{ .82,  .898,  .941}96.6 & \cellcolor[rgb]{ .937,  .541,  .384}90.1 & \cellcolor[rgb]{ .129,  .4,  .675}\textcolor[rgb]{ 1,  1,  1}{72.5} & \cellcolor[rgb]{ .404,  .663,  .812}67.2 & 99.3  & \cellcolor[rgb]{ .937,  .541,  .384}67.2 & \cellcolor[rgb]{ .404,  .663,  .812}80.6 & 99.3 \bigstrut[b]\\
\hline
\end{tabular}%

		\normalsize
		\caption{Overall payoffs to animal $A$ in parameter regime II when it pursues the strategy in the row and animal $B$ pursues the strategy in the column. Our parameter values are $c_W = 0.1$, $c_L = 0.6$, $\gamma = 0.995$, and $k = 1$.
			All other details are identical to those for Table \ref{T:RegimeIGridOfOutputs}.
			\label{T:RegimeIIGridOfOutputs}
		}
	\end{table}

	\begin{table}[ht]
		\centering
		\footnotesize 
\setlength\tabcolsep{3pt}
\begin{tabular}{|l|llllllllllllll|}
\hline
      & A     & B     & C     & D     & E     & F     & G     & H     & I     & J     & K     & L     & M     & N \bigstrut\\
\hline
(A) AllH & $-$29.7 & \cellcolor[rgb]{ .992,  .859,  .78}27.2\phantom{$-$} & \cellcolor[rgb]{ .937,  .541,  .384}84.5\phantom{$-$} & \cellcolor[rgb]{ .698,  .094,  .169}\textcolor[rgb]{ 1,  1,  1}{141.5} & \cellcolor[rgb]{ .698,  .094,  .169}\textcolor[rgb]{ 1,  1,  1}{198.7} & \cellcolor[rgb]{ .129,  .4,  .675}\textcolor[rgb]{ 1,  1,  1}{$-$28.7} & \cellcolor[rgb]{ .129,  .4,  .675}\textcolor[rgb]{ 1,  1,  1}{$-$28.7} & \cellcolor[rgb]{ .404,  .663,  .812}84.5\phantom{$-$} & \cellcolor[rgb]{ .129,  .4,  .675}\textcolor[rgb]{ 1,  1,  1}{25.8}\phantom{$-$} & \cellcolor[rgb]{ .129,  .4,  .675}\textcolor[rgb]{ 1,  1,  1}{33.0}\phantom{$-$} & \cellcolor[rgb]{ .129,  .4,  .675}\textcolor[rgb]{ 1,  1,  1}{25.1} & \cellcolor[rgb]{ .129,  .4,  .675}\textcolor[rgb]{ 1,  1,  1}{50.1}\phantom{$-$} & \cellcolor[rgb]{ .129,  .4,  .675}\textcolor[rgb]{ 1,  1,  1}{$-$19.0} & \cellcolor[rgb]{ .129,  .4,  .675}\textcolor[rgb]{ 1,  1,  1}{$-$18.7} \bigstrut[t]\\
(B) 25\%D & \cellcolor[rgb]{ .937,  .541,  .384}$-$22.2 & 26.8  & \cellcolor[rgb]{ .937,  .541,  .384}75.8 & \cellcolor[rgb]{ .698,  .094,  .169}\textcolor[rgb]{ 1,  1,  1}{124.8} & \cellcolor[rgb]{ .698,  .094,  .169}\textcolor[rgb]{ 1,  1,  1}{173.8} & \cellcolor[rgb]{ .129,  .4,  .675}\textcolor[rgb]{ 1,  1,  1}{27.4} & \cellcolor[rgb]{ .129,  .4,  .675}\textcolor[rgb]{ 1,  1,  1}{$-$21.2} & \cellcolor[rgb]{ .129,  .4,  .675}\textcolor[rgb]{ 1,  1,  1}{75.7} & \cellcolor[rgb]{ .129,  .4,  .675}\textcolor[rgb]{ 1,  1,  1}{27.4} & \cellcolor[rgb]{ .129,  .4,  .675}\textcolor[rgb]{ 1,  1,  1}{37.2} & \cellcolor[rgb]{ .129,  .4,  .675}\textcolor[rgb]{ 1,  1,  1}{30.1} & \cellcolor[rgb]{ .129,  .4,  .675}\textcolor[rgb]{ 1,  1,  1}{68.0} & \cellcolor[rgb]{ .129,  .4,  .675}\textcolor[rgb]{ 1,  1,  1}{5.6} & \cellcolor[rgb]{ .129,  .4,  .675}\textcolor[rgb]{ 1,  1,  1}{22.3} \\
(C) 50\%D & \cellcolor[rgb]{ .937,  .541,  .384}$-$14.9 & \cellcolor[rgb]{ .82,  .898,  .941}26.1 & 67.0  & \cellcolor[rgb]{ .937,  .541,  .384}108.0 & \cellcolor[rgb]{ .698,  .094,  .169}\textcolor[rgb]{ 1,  1,  1}{149.0} & \cellcolor[rgb]{ .129,  .4,  .675}\textcolor[rgb]{ 1,  1,  1}{67.5} & \cellcolor[rgb]{ .129,  .4,  .675}\textcolor[rgb]{ 1,  1,  1}{$-$13.3} & \cellcolor[rgb]{ .129,  .4,  .675}\textcolor[rgb]{ 1,  1,  1}{67.1} & \cellcolor[rgb]{ .404,  .663,  .812}29.3 & \cellcolor[rgb]{ .129,  .4,  .675}\textcolor[rgb]{ 1,  1,  1}{41.3} & \cellcolor[rgb]{ .129,  .4,  .675}\textcolor[rgb]{ 1,  1,  1}{35.4} & \cellcolor[rgb]{ .129,  .4,  .675}\textcolor[rgb]{ 1,  1,  1}{79.3} & \cellcolor[rgb]{ .404,  .663,  .812}26.2 & \cellcolor[rgb]{ .129,  .4,  .675}\textcolor[rgb]{ 1,  1,  1}{52.5} \\
(D) 75\%D & \cellcolor[rgb]{ .698,  .094,  .169}\textcolor[rgb]{ 1,  1,  1}{$-$7.5} & \cellcolor[rgb]{ .82,  .898,  .941}25.5 & \cellcolor[rgb]{ .404,  .663,  .812}58.4 & 91.3\phantom{$-$}  & \cellcolor[rgb]{ .698,  .094,  .169}\textcolor[rgb]{ 1,  1,  1}{124.2} & \cellcolor[rgb]{ .404,  .663,  .812}91.4 & \cellcolor[rgb]{ .129,  .4,  .675}\textcolor[rgb]{ 1,  1,  1}{$-$4.9} & \cellcolor[rgb]{ .129,  .4,  .675}\textcolor[rgb]{ 1,  1,  1}{58.4} & \cellcolor[rgb]{ .404,  .663,  .812}31.3 & \cellcolor[rgb]{ .129,  .4,  .675}\textcolor[rgb]{ 1,  1,  1}{45.6} & \cellcolor[rgb]{ .129,  .4,  .675}\textcolor[rgb]{ 1,  1,  1}{40.6} & \cellcolor[rgb]{ .404,  .663,  .812}83.9 & \cellcolor[rgb]{ .937,  .541,  .384}43.2 & \cellcolor[rgb]{ .129,  .4,  .675}\textcolor[rgb]{ 1,  1,  1}{72.3} \\
(E) AllD & \cellcolor[rgb]{ .698,  .094,  .169}\textcolor[rgb]{ 1,  1,  1}{0.0} & \cellcolor[rgb]{ .82,  .898,  .941}24.9 & \cellcolor[rgb]{ .404,  .663,  .812}49.6 & \cellcolor[rgb]{ .404,  .663,  .812}74.5 & 99.3\phantom{$-$}  & 99.3  & 99.3  & \cellcolor[rgb]{ .129,  .4,  .675}\textcolor[rgb]{ 1,  1,  1}{49.8} & \cellcolor[rgb]{ .992,  .859,  .78}49.8 & \cellcolor[rgb]{ .129,  .4,  .675}\textcolor[rgb]{ 1,  1,  1}{49.7} & \cellcolor[rgb]{ .129,  .4,  .675}\textcolor[rgb]{ 1,  1,  1}{45.8} & \cellcolor[rgb]{ .404,  .663,  .812}81.9 & \cellcolor[rgb]{ .698,  .094,  .169}\textcolor[rgb]{ 1,  1,  1}{56.4} & \cellcolor[rgb]{ .404,  .663,  .812}81.9 \\
(F) TfT & $-$29.7 & 26.6  & 66.9  & 91.2  & 99.3  & 99.3  & 99.3  & \cellcolor[rgb]{ .129,  .4,  .675}\textcolor[rgb]{ 1,  1,  1}{69.7} & \cellcolor[rgb]{ .404,  .663,  .812}31.2 & \cellcolor[rgb]{ .129,  .4,  .675}\textcolor[rgb]{ 1,  1,  1}{$-$16.3} & \cellcolor[rgb]{ .129,  .4,  .675}\textcolor[rgb]{ 1,  1,  1}{$-$17.9} & \cellcolor[rgb]{ .129,  .4,  .675}\textcolor[rgb]{ 1,  1,  1}{47.1} & \cellcolor[rgb]{ .129,  .4,  .675}\textcolor[rgb]{ 1,  1,  1}{$-$6.4} & \cellcolor[rgb]{ .129,  .4,  .675}\textcolor[rgb]{ 1,  1,  1}{47.0} \\
(G) Grim & $-$29.6 & \cellcolor[rgb]{ .992,  .859,  .78}27.3 & \cellcolor[rgb]{ .937,  .541,  .384}84.1 & \cellcolor[rgb]{ .698,  .094,  .169}\textcolor[rgb]{ 1,  1,  1}{140.4} & 99.3  & 99.3  & 99.3  & \cellcolor[rgb]{ .404,  .663,  .812}86.4 & \cellcolor[rgb]{ .404,  .663,  .812}37.6 & \cellcolor[rgb]{ .129,  .4,  .675}\textcolor[rgb]{ 1,  1,  1}{$-$16.2} & \cellcolor[rgb]{ .129,  .4,  .675}\textcolor[rgb]{ 1,  1,  1}{$-$17.7} & \cellcolor[rgb]{ .129,  .4,  .675}\textcolor[rgb]{ 1,  1,  1}{47.0} & \cellcolor[rgb]{ .129,  .4,  .675}\textcolor[rgb]{ 1,  1,  1}{$-$6.6} & \cellcolor[rgb]{ .129,  .4,  .675}\textcolor[rgb]{ 1,  1,  1}{47.1} \\
(H) Pav & \cellcolor[rgb]{ .937,  .541,  .384}$-$14.8 & \cellcolor[rgb]{ .82,  .898,  .941}26.1 & 67.1  & \cellcolor[rgb]{ .937,  .541,  .384}107.9 & \cellcolor[rgb]{ .698,  .094,  .169}\textcolor[rgb]{ 1,  1,  1}{149.2} & \cellcolor[rgb]{ .129,  .4,  .675}\textcolor[rgb]{ 1,  1,  1}{70.2} & \cellcolor[rgb]{ .129,  .4,  .675}\textcolor[rgb]{ 1,  1,  1}{6.0} & 96.4  & \cellcolor[rgb]{ .698,  .094,  .169}\textcolor[rgb]{ 1,  1,  1}{80.1} & \cellcolor[rgb]{ .129,  .4,  .675}\textcolor[rgb]{ 1,  1,  1}{41.4} & \cellcolor[rgb]{ .129,  .4,  .675}\textcolor[rgb]{ 1,  1,  1}{35.3} & \cellcolor[rgb]{ .404,  .663,  .812}80.3 & \cellcolor[rgb]{ .404,  .663,  .812}26.9 & \cellcolor[rgb]{ .129,  .4,  .675}\textcolor[rgb]{ 1,  1,  1}{54.2} \\
(I) MPav & \cellcolor[rgb]{ .698,  .094,  .169}\textcolor[rgb]{ 1,  1,  1}{$-$5.0} & \cellcolor[rgb]{ .937,  .541,  .384}38.2 & \cellcolor[rgb]{ .937,  .541,  .384}81.0 & \cellcolor[rgb]{ .698,  .094,  .169}\textcolor[rgb]{ 1,  1,  1}{123.7} & \cellcolor[rgb]{ .698,  .094,  .169}\textcolor[rgb]{ 1,  1,  1}{148.8} & \cellcolor[rgb]{ .129,  .4,  .675}\textcolor[rgb]{ 1,  1,  1}{64.7} & \cellcolor[rgb]{ .129,  .4,  .675}\textcolor[rgb]{ 1,  1,  1}{13.9} & \cellcolor[rgb]{ .937,  .541,  .384}101.0 & 48.7  & \cellcolor[rgb]{ .129,  .4,  .675}\textcolor[rgb]{ 1,  1,  1}{37.6} & \cellcolor[rgb]{ .129,  .4,  .675}\textcolor[rgb]{ 1,  1,  1}{31.9} & \cellcolor[rgb]{ .129,  .4,  .675}\textcolor[rgb]{ 1,  1,  1}{76.0} & \cellcolor[rgb]{ .404,  .663,  .812}18.8 & \cellcolor[rgb]{ .129,  .4,  .675}\textcolor[rgb]{ 1,  1,  1}{45.2} \\
(J) Bul, N, PDT \, & \cellcolor[rgb]{ .698,  .094,  .169}\textcolor[rgb]{ 1,  1,  1}{36.5} & \cellcolor[rgb]{ .698,  .094,  .169}\textcolor[rgb]{ 1,  1,  1}{64.6} & \cellcolor[rgb]{ .698,  .094,  .169}\textcolor[rgb]{ 1,  1,  1}{92.7} & \cellcolor[rgb]{ .698,  .094,  .169}\textcolor[rgb]{ 1,  1,  1}{120.9} & \cellcolor[rgb]{ .698,  .094,  .169}\textcolor[rgb]{ 1,  1,  1}{149.0} & \cellcolor[rgb]{ .404,  .663,  .812}86.5 & \cellcolor[rgb]{ .404,  .663,  .812}86.4 & \cellcolor[rgb]{ .404,  .663,  .812}92.8 & \cellcolor[rgb]{ .698,  .094,  .169}\textcolor[rgb]{ 1,  1,  1}{86.8} & 99.3  & \cellcolor[rgb]{ .992,  .859,  .78}90.2 & \cellcolor[rgb]{ .698,  .094,  .169}\textcolor[rgb]{ 1,  1,  1}{131.5} & \cellcolor[rgb]{ .698,  .094,  .169}\textcolor[rgb]{ 1,  1,  1}{53.8} & \cellcolor[rgb]{ .129,  .4,  .675}\textcolor[rgb]{ 1,  1,  1}{79.4} \\
(K) MN & \cellcolor[rgb]{ .698,  .094,  .169}\textcolor[rgb]{ 1,  1,  1}{34.5} & \cellcolor[rgb]{ .698,  .094,  .169}\textcolor[rgb]{ 1,  1,  1}{64.2} & \cellcolor[rgb]{ .698,  .094,  .169}\textcolor[rgb]{ 1,  1,  1}{93.6} & \cellcolor[rgb]{ .698,  .094,  .169}\textcolor[rgb]{ 1,  1,  1}{123.3} & \cellcolor[rgb]{ .698,  .094,  .169}\textcolor[rgb]{ 1,  1,  1}{152.8} & \cellcolor[rgb]{ .404,  .663,  .812}83.7 & \cellcolor[rgb]{ .129,  .4,  .675}\textcolor[rgb]{ 1,  1,  1}{77.7} & \cellcolor[rgb]{ .404,  .663,  .812}93.7 & \cellcolor[rgb]{ .698,  .094,  .169}\textcolor[rgb]{ 1,  1,  1}{85.5} & \cellcolor[rgb]{ .404,  .663,  .812}96.4 & 88.7  & \cellcolor[rgb]{ .698,  .094,  .169}\textcolor[rgb]{ 1,  1,  1}{135.4} & \cellcolor[rgb]{ .937,  .541,  .384}51.2 & \cellcolor[rgb]{ .129,  .4,  .675}\textcolor[rgb]{ 1,  1,  1}{76.6} \\
(L) Slf & \cellcolor[rgb]{ .698,  .094,  .169}\textcolor[rgb]{ 1,  1,  1}{37.4} & \cellcolor[rgb]{ .698,  .094,  .169}\textcolor[rgb]{ 1,  1,  1}{59.6} & \cellcolor[rgb]{ .937,  .541,  .384}80.2 & \cellcolor[rgb]{ .937,  .541,  .384}99.3 & \cellcolor[rgb]{ .937,  .541,  .384}116.8 & \cellcolor[rgb]{ .937,  .541,  .384}106.4 & \cellcolor[rgb]{ .937,  .541,  .384}106.4 & \cellcolor[rgb]{ .404,  .663,  .812}80.5 & \cellcolor[rgb]{ .698,  .094,  .169}\textcolor[rgb]{ 1,  1,  1}{79.1} & \cellcolor[rgb]{ .129,  .4,  .675}\textcolor[rgb]{ 1,  1,  1}{67.2} & \cellcolor[rgb]{ .129,  .4,  .675}\textcolor[rgb]{ 1,  1,  1}{63.3} & 99.3  & \cellcolor[rgb]{ .698,  .094,  .169}\textcolor[rgb]{ 1,  1,  1}{73.9} & 99.3 \\
(M) SDT & \cellcolor[rgb]{ .698,  .094,  .169}\textcolor[rgb]{ 1,  1,  1}{5.2} & \cellcolor[rgb]{ .698,  .094,  .169}\textcolor[rgb]{ 1,  1,  1}{52.9} & \cellcolor[rgb]{ .698,  .094,  .169}\textcolor[rgb]{ 1,  1,  1}{91.4} & \cellcolor[rgb]{ .698,  .094,  .169}\textcolor[rgb]{ 1,  1,  1}{121.3} & \cellcolor[rgb]{ .698,  .094,  .169}\textcolor[rgb]{ 1,  1,  1}{142.2} & \cellcolor[rgb]{ .404,  .663,  .812}94.1 & \cellcolor[rgb]{ .404,  .663,  .812}94.1 & \cellcolor[rgb]{ .404,  .663,  .812}93.0 & \cellcolor[rgb]{ .698,  .094,  .169}\textcolor[rgb]{ 1,  1,  1}{74.4} & \cellcolor[rgb]{ .129,  .4,  .675}\textcolor[rgb]{ 1,  1,  1}{61.4} & \cellcolor[rgb]{ .129,  .4,  .675}\textcolor[rgb]{ 1,  1,  1}{59.8} & \cellcolor[rgb]{ .698,  .094,  .169}\textcolor[rgb]{ 1,  1,  1}{124.8} & 33.4  & \cellcolor[rgb]{ .404,  .663,  .812}87.0 \\
(N) InT & \cellcolor[rgb]{ .698,  .094,  .169}\textcolor[rgb]{ 1,  1,  1}{5.3} & \cellcolor[rgb]{ .698,  .094,  .169}\textcolor[rgb]{ 1,  1,  1}{48.8} & \cellcolor[rgb]{ .937,  .541,  .384}81.9 & \cellcolor[rgb]{ .937,  .541,  .384}104.6 & \cellcolor[rgb]{ .937,  .541,  .384}116.8 & \cellcolor[rgb]{ .937,  .541,  .384}106.3 & \cellcolor[rgb]{ .937,  .541,  .384}106.4 & \cellcolor[rgb]{ .404,  .663,  .812}83.8 & \cellcolor[rgb]{ .937,  .541,  .384}67.7 & \cellcolor[rgb]{ .129,  .4,  .675}\textcolor[rgb]{ 1,  1,  1}{36.1} & \cellcolor[rgb]{ .129,  .4,  .675}\textcolor[rgb]{ 1,  1,  1}{34.3} & 99.3  & \cellcolor[rgb]{ .937,  .541,  .384}45.8 & 99.3 \bigstrut[b]\\
\hline
\end{tabular}%

		\normalsize
		\caption{Overall payoffs to animal $A$ in parameter regime III when it pursues the strategy in the row and animal $B$ pursues the strategy in the column. Our parameter values are $c_W = 0.1$, $c_L = 1.2$, $\gamma = 0.995$, and $k = 1$.
			All other details are identical to those for Table \ref{T:RegimeIGridOfOutputs}.			
			\label{T:RegimeIIIGridOfOutputs}
		}
	\end{table}
\end{landscape}


\section{Pair behaviours associated with the PD TfT strategy}
\label{S:PairBehaviour}

Suppose that animals $A$ and $B$ both pursue PD TfT strategies. For fixed values of $c_L$ and $c_W$, animal $A$'s substrategy (Always Dove, Tit for Tat, or Always Hawk) depends on the value of $p_A$. 
To determine the parameter values that are associated with different pair behaviours,
we seek critical values of $p_A$ that separate the different substrategies of PD TfT. We define $\tilde{p}_1$ and $\tilde{p}_2$ such that animal $A$ pursues Always Dove when $0 \leq p_A < \tilde{p}_1$, Tit for Tat when $\tilde{p}_1 < p_A < \tilde{p}_2$, and Always Hawk when $\tilde{p}_2 < p_A \leq 1$. From the definition of PD TfT in Table \ref{T:PlausibleStrategies} and the boundaries between stage-game classifications (see Table \ref{T:StageGameRegimes}), we calculate $\tilde{p}_1$ and $\tilde{p}_2$ in the different parameter regimes and obtain the results in Table \ref{T:CriticalP}.

\begin{table}[ht]
	\centering 
	\begin{tabular}{|c|c|c|}
		\hline
		\multirow{2}[4]{*}{Parameter regimes} & \multicolumn{2}{c|}{Critical values of $p_A$} \bigstrut\\
		\cline{2-3}      & $\tilde{p}_1$ & $\tilde{p}_2$ \bigstrut\\
		\hline
		I     &   $ c_L/(1 - c_W + c_L)$    &  $(\frac{1}{2} + c_L)/(1 - c_W + c_L)$ \bigstrut\\ 
		\hline
		II and IV &    $ c_L/(1 - c_W + c_L)$   &  $ (1 - c_W)/(1 - c_W + c_L)$ \bigstrut\\
		\hline
		III, V, and VI &    $\tfrac{1}{2}$   &  $\tfrac{1}{2}$ \bigstrut\\
		\hline
	\end{tabular}
	\caption{Critical values, $\tilde{p}_1$ and $\tilde{p}_2$, of $p_A$ in the PD TfT
		strategy. Animal $A$ pursues Always Dove when $0 \leq p_A < \tilde{p}_1$, Tit for Tat when $\tilde{p}_1 < p_A < \tilde{p}_2$, and Always Hawk when $\tilde{p}_2 < p_A \leq 1$. Tit for Tat is not used in parameter regimes III, V, and VI, so $\tilde{p}_1 = \tilde{p}_2$ in those regimes. 
	} 
	\label{T:CriticalP}
\end{table}

If both animals in a contest pursue a PD TfT strategy, there are three different pair behaviours (depending on the values of $\mu_A$ and $\mu_B$); we show these using coloured shading in Figure \ref{F:muValuesAndClassification}. If the stage game is PD, both animals use a TfT substrategy, and we observe that each animal plays Dove in every round. If the stage game is PD--Deadlock or Deadlock--PD, one animal uses a TfT substrategy, and the other uses an Always Hawk substrategy. Consequently, both animals play Hawk in every round except for the first one. For any other stage game, the animal with the larger RHP uses an Always Hawk substrategy, and the animal with the smaller RHP uses Always Dove. This corresponds to a dominance relationship.

All three of the above possible pair behaviours can occur in parameter regime I. 
If there is a sufficiently large difference in the RHPs of the two animals (specifically, if $|p_A - \frac{1}{2}| > \frac{1 - c_W - c_L}{2(1-c_W + c_L)}$), fighting is too costly for the animal with the smaller RHP, so there is a dominance relationship.
If there is a moderate difference in the RHPs of the two animals (specifically, if $\frac{1 - c_W - c_L}{2(1-c_W + c_L)} < |p_A - \frac{1}{2}| <  \frac{c_W + c_L}{2(1-c_W + c_L)}$), fighting is not too costly, and the two animals fight in every round. The weaker animal pursues a strategy of Tit for Tat, but this is not a sufficient threat to prevent the stronger animal from playing Hawk in every round, leading to continual fighting. If there is a sufficiently small difference in the RHPs of the two animals (specifically, if $|p_A - \frac{1}{2}| <  \frac{c_W + c_L}{2(1-c_W + c_L)}$), sharing resources gives a better payoff for the stronger animal than continual fighting. Therefore, Tit for Tat is a credible threat, and both animals play Dove in every round to avoid the cost of fighting. 

In parameter regime II (and also in regime IV), the possible pair behaviours are dominance relationships (when $|p_A - \frac{1}{2}| > \frac{1 - c_W - c_L}{2(1-c_W + c_L)}$) and resource sharing (when $|p_A - \frac{1}{2}| <   \frac{1 - c_W - c_L}{2(1-c_W + c_L)}$). Sharing, in which both animals pursue a Tit for Tat substrategy, occurs if and only if the expected payoff from a fight is positive for both animals. In this situation, the animal with the smaller RHP expects to do better from a fight than it would from playing Dove against its opponent's Hawk, and the animal with the larger RHP expects to do better from sharing than it would from a fight.
Therefore, the animal with the smaller RHP can credibly threaten the animal with the larger RHP, and Tit for Tat becomes an effective substrategy for both animals. 

In parameter regime III (and also in regimes V and VI), the only possible pair behaviour is a dominance relationship.
The animal with the larger RHP always plays Hawk, and the animal with the smaller RHP always plays Dove.
Because fighting is costly (especially for the loser) in parameter regime III, dominance relationships are favored, even when the difference between the animals' RHPs is arbitrarily small.

We see from the above results that the qualitative types of pair behaviours in our observations differ across the parameter regimes. In parameter regime I, we observe all three pair behaviours; in parameter regimes II and IV, we observe only dominance relationships and resource sharing; in parameter regimes III, V, and VI, we observe only dominance relationships.
From Figure \ref{F:StageGameRegimes}, it follows that $c_W + c_L$ (i.e., the sum of the fighting cost for the two animals) is a key parameter for determining possible pair behaviours. If $c_W + c_L < \frac{1}{2}$, a dominance relationship, continual fights, or sharing of resources can occur; if $\frac{1}{2} < c_W + c_L < 1$, sharing of resources or dominance relationships can occur; and if $c_W + c_L > 1$, there are only dominance relationships.


\section{Incorporation of Tit for Tat into learning strategies}
\label{S:TfTTransitions}

The first is a delay in adopting Tit for Tat. We specify that any animal that shifts from Always Hawk to Tit for Tat first plays Dove for two rounds before beginning the Tit for Tat process. To understand why this is important, suppose that animal $B$ pursues a Tit for Tat strategy and that animal $A$ changes its substrategy from Always Hawk to Tit for Tat after round $m$. Because animal $A$ has been playing Always Hawk up to this point and animal $B$ is imitating this choice in its Tit for Tat strategy, animal $B$'s move during round $m$ is Hawk. If animal $A$ immediately switches to Tit for Tat for round $m+1$, the two animals continue to play Hawk during every move.
If animal $A$ plays Dove during round $m+1$ and then starts copying animal $B$'s previous move starting from round $m+2$, the two animals will alternate Dove--Hawk and Hawk--Dove in all rounds starting from $m+1$. The only way to create the possibility of Dove--Dove interactions in all future rounds is for animal $A$ to play Dove for two rounds before pursuing a Tit for Tat substrategy.

The second feature is hysteresis, which prevents overly frequent changes in substrategies. For mean-based and median-based strategies, we specify two priors. One (the main prior, with parameters $\alpha_0$ and $\beta_0$) determines whether to change to a less aggressive substrategy, and the other (the secondary prior, with parameters $\bar{\alpha}_0$ and $\bar{\beta}_0$) determines whether to change to a more aggressive substrategy.
We require the secondary prior to be less optimistic about $p_A$ than the main prior by demanding that $\bar{\alpha}_0 \leq \alpha_0$ and $\bar{\beta}_0 \geq \beta_0$.

Suppose that animal $A$ pursues a mean-based learning strategy in which the secondary prior is different from the main prior. We use the main prior to decide the substrategy for the animal's first move of the contest. If $\frac{\alpha_0}{\alpha_0 + \beta_0} \geq \tilde{p}_2$, the animal's substrategy in the first round is Always Hawk. As the contest progresses, if the mean $\frac{\alpha}{\alpha+\beta}$ of the main distribution goes from above $\tilde{p}_2$ to below $\tilde{p}_2$, then $\substratDoveProbs$ changes from $\substratDoveProbs_\text{H}$ to $\substratDoveProbs_\text{TfT}$. The main prior thereby determines whether to change from a more-aggressive substrategy (Always Hawk) to a less-aggressive substrategy (Tit for Tat).
If $\frac{\alpha}{\alpha+\beta}$ subsequently goes from below $\tilde{p}_2$ to above $\tilde{p}_2$, an animal does not immediately change substrategy again. If the true value of $p_A$ is near $\tilde{p}_2$, there may be frequent changes between Always Hawk and Tit for Tat as $\frac{\alpha}{\alpha+\beta}$ fluctuates around $\tilde{p}_2$. Because an animal that changes from Always Hawk to Tit for Tat plays Dove for at least two rounds, it is suboptimal for it to make frequent changes in its substrategy. 
Instead, it changes from $\substratDoveProbs_\text{TfT}$ back to $\substratDoveProbs_\text{H}$ only if the mean $\frac{\bar{\alpha}}{\bar{\alpha}+\bar{\beta}}$ of the secondary distribution increases above $\tilde{p}_2$.
Because $\bar{\alpha}_0 \leq \alpha_0$ and $\bar{\beta}_0 \geq \beta_0$ (where equality cannot occur simultaneously in these two inequalities when the main and secondary priors differ from each other), it follows that $\frac{\bar{\alpha}}{\bar{\alpha}+\bar{\beta}} < \frac{\alpha}{\alpha+\beta}$, so $\frac{\bar{\alpha}}{\bar{\alpha}+\bar{\beta}}$ increases above $\tilde{p}_2$ only after $\frac{\alpha}{\alpha+\beta}$ increases above $\tilde{p}_2$. 

We also introduce hysteresis for high-quantile strategies, but now we do so by specifying a main quantile and a secondary quantile, rather than by specifying a main prior and a secondary prior. We require the secondary quantile to be lower than the main quantile.
In a quantile-based strategy, an animal changes from Always Hawk to Tit for Tat if the main quantile of $\mathrm{Beta}(\alpha,\beta)$ goes from above $\tilde{p}_2$ to below $\tilde{p}_2$; it changes from Tit for Tat to Always Hawk if the secondary quantile of $\mathrm{Beta}(\alpha,\beta)$ goes from below $\tilde{p}_2$ to above $\tilde{p}_2$. Similarly, an animal changes from Tit for Tat to Always Dove if the main quantile of $\mathrm{Beta}(\alpha,\beta)$ goes from above $\tilde{p}_1$ to below it.


\section[Results for animals with knowledge of pA]{Tournament results for animals with no knowledge of $p_A$}
\label{S:ResultsLearning}

In Table \ref{T:LearningStratsTable}, we list the 25 different strategies (which we label from `A' to `Y') in our tournament for animals with no knowledge of $p_A$. We show the results of our tournaments in Tables \ref{T:RegimeILearningColourOnly} (for parameter regime I), \ref{T:RegimeIILearningColourOnly} (for regime II), and \ref{T:RegimeIIILearningColourOnly} (for regime III). We colour the cells of Tables \ref{T:RegimeILearningColourOnly}, \ref{T:RegimeIILearningColourOnly}, and \ref{T:RegimeIIILearningColourOnly} to indicate whether a mutant who pursues a strategy that is indicated by the row can invade a population of animals who pursue the strategy that is indicated by the column. This colouring system is identical to the colouring that we described in Appendix \ref{S:ResultsInformed} for Tables \ref{T:RegimeIGridOfOutputs}, \ref{T:RegimeIIGridOfOutputs}, and \ref{T:RegimeIIIGridOfOutputs}.
	
Because of space limitations (and in contrast to Tables \ref{T:RegimeIGridOfOutputs}, \ref{T:RegimeIIGridOfOutputs}, and \ref{T:RegimeIIIGridOfOutputs}), we do not give numerical results for the mean payoff to animal $A$ when it pursues the strategy in the row and animal $B$ pursues the strategy in the column. These results are available at \url{https://bitbucket.org/CameronLHall/dominancesharingassessmentmatlab/}.
		
Strategies N--U are examples of mean-based and median-based learning strategies. As we described in Appendix \ref{S:TfTTransitions}, we introduce hysteresis to the transitions between Tit for Tat and Always Hawk using a `secondary prior' for deciding when to change from a less-aggressive substrategy to a more-aggressive one. For all of these strategies, we use a flat prior as the secondary prior (which entails that $\bar{\alpha}_0 = \bar{\beta}_0 = 1$).
	
Strategies V--Y are high-quantile strategies. For these strategies, we introduce hysteresis using a `secondary quantile' for decisions that concern changing from a less-aggressive substrategy to a more-aggressive one. For strategy V, we use a secondary quantile of 0.7; for W, we use 0.8; for X, we use 0.9; for Y, we use 0.95. To determine appropriate values for these secondary quantiles, we ran a series of simulations in which animal $A$ used a high-quantile strategy, animal $B$ used an Always Hawk strategy, and $p_A$ was approximately equal to $\tilde{p}_2$. We chose the secondary quantiles to be sufficiently close to the primary quantiles to enable animals to change substrategy from Always Hawk to Tit for Tat and back again, but sufficiently far away from the primary quantiles to ensure that it was rare for an animal to make three or more changes in substrategy (e.g., from Always Hawk to Tit for Tat to Always Hawk to Tit for Tat).

\begin{table}[ht]
	\centering
	\scriptsize
\begin{tabular}{|c|l|l|}
\hline
Label & \multicolumn{1}{c|}{Strategy type} & \multicolumn{1}{c|}{Description/details of prior} \bigstrut\\
\hline
A     & Simple & Always Hawk \bigstrut\\
\hline
B     & Simple & 25\% Dove \bigstrut\\
\hline
C     & Simple & 50\% Dove \bigstrut\\
\hline
D     & Simple & 75\% Dove \bigstrut\\
\hline
E     & Simple & Always Dove \bigstrut\\
\hline
F     & Simple & Tit for Tat \bigstrut\\
\hline
G     & Simple & Grim \bigstrut\\
\hline
\end{tabular}%

	\normalsize
	\caption{Our 25 strategies in our tournaments of learning strategies. The first column indicates the label for the strategy (see Tables \ref{T:RegimeILearningColourOnly}--\ref{T:RegimeIIILearningColourOnly}), the second column indicates the strategy type, and the third column gives some further details about the strategy.}
	\label{T:LearningStratsTable}
	
\end{table}

\begin{landscape}
	\begin{table}[ht]
		\centering
		\ssmall
\begin{tabular}{|cc|lllllllll|llll|llll|llll|llll|}
\hline
      &       & \multicolumn{9}{c|}{Simple}                                           & \multicolumn{4}{c|}{Weighted} & \multicolumn{4}{c|}{Mean}     & \multicolumn{4}{c|}{Median}   & \multicolumn{4}{c|}{Quantile} \bigstrut[t]\\
      &       & \multicolumn{1}{c}{A} & \multicolumn{1}{c}{B} & \multicolumn{1}{c}{C} & \multicolumn{1}{c}{D} & \multicolumn{1}{c}{E} & \multicolumn{1}{c}{F} & \multicolumn{1}{c}{G} & \multicolumn{1}{c}{H} & \multicolumn{1}{c|}{I} & \multicolumn{1}{c}{J} & \multicolumn{1}{c}{K} & \multicolumn{1}{c}{L} & \multicolumn{1}{c|}{M} & \multicolumn{1}{c}{N} & \multicolumn{1}{c}{O} & \multicolumn{1}{c}{P} & \multicolumn{1}{c|}{Q} & \multicolumn{1}{c}{R} & \multicolumn{1}{c}{S} & \multicolumn{1}{c}{T} & \multicolumn{1}{c|}{U} & \multicolumn{1}{c}{V} & \multicolumn{1}{c}{W} & \multicolumn{1}{c}{X} & \multicolumn{1}{c|}{Y} \bigstrut[b]\\
\hline
\end{tabular}%

		\normalsize
		\caption{Overall payoffs to animal $A$ in parameter regime I when it		
			pursues the strategy in the row and animal $B$ pursues the strategy in the column.
			The letters in the rows and columns correspond to the strategies in Table \ref{T:LearningStratsTable}.
			The parameter values are $c_W = 0.1$, $c_L = 0.2$, $\gamma = 0.995$, and $k = 1$.
			We colour cells according to $Q(T,S) = [E(T,S) - E(S,S)]/[E(\text{AllH},\text{AllD})]$, where $T$ is the strategy that is pursued by animal $A$ (indicated by the row) and $S$ is the strategy that is pursued by animal $B$ (indicated by the column).
			Cells are red when $Q(T,S) > 10^{-3}$, blue when $Q(T,S) < -10^{-3}$, and white when $|Q(T,S)| \leq 10^{-3}$.
			Cells are dark when
			$|Q(T,S)| > 10^{-1}$, of a medium shade when
			$10^{-2} <  |Q(T,S)| \leq 10^{-1}$, and light when
			$10^{-3} < |Q(T,S)| < 10^{-2}$.}
		\label{T:RegimeILearningColourOnly}
	\end{table}

	\begin{table}[ht]
		\centering
		\ssmall
\begin{tabular}{|cc|lllllllll|llll|llll|llll|llll|}
\hline
      &       & \multicolumn{9}{c|}{Simple}                                           & \multicolumn{4}{c|}{Weighted} & \multicolumn{4}{c|}{Mean}     & \multicolumn{4}{c|}{Median}   & \multicolumn{4}{c|}{Quantile} \bigstrut[t]\\
      &       & \multicolumn{1}{c}{A} & \multicolumn{1}{c}{B} & \multicolumn{1}{c}{C} & \multicolumn{1}{c}{D} & \multicolumn{1}{c}{E} & \multicolumn{1}{c}{F} & \multicolumn{1}{c}{G} & \multicolumn{1}{c}{H} & \multicolumn{1}{c|}{I} & \multicolumn{1}{c}{J} & \multicolumn{1}{c}{K} & \multicolumn{1}{c}{L} & \multicolumn{1}{c|}{M} & \multicolumn{1}{c}{N} & \multicolumn{1}{c}{O} & \multicolumn{1}{c}{P} & \multicolumn{1}{c|}{Q} & \multicolumn{1}{c}{R} & \multicolumn{1}{c}{S} & \multicolumn{1}{c}{T} & \multicolumn{1}{c|}{U} & \multicolumn{1}{c}{V} & \multicolumn{1}{c}{W} & \multicolumn{1}{c}{X} & \multicolumn{1}{c|}{Y} \bigstrut[b]\\
\hline
\end{tabular}%

		\normalsize
		\caption{Overall payoffs to animal $A$ in parameter regime II when it pursues the strategy in the row and animal $B$ pursues the strategy in the column. The letters in the rows and columns correspond to the strategies in Table \ref{T:LearningStratsTable}.
			The parameter values are $c_W = 0.1$, $c_L = 0.6$, $\gamma = 0.995$, and $k = 1$.
			We colour cells as in Table \ref{T:RegimeILearningColourOnly}.}
		\label{T:RegimeIILearningColourOnly}
	\end{table}

	\begin{table}[ht]
		\centering
		\ssmall
\begin{tabular}{|cc|ccccccccc|cccc|cccc|cccc|cccc|}
\hline
      &       & \multicolumn{9}{c|}{Simple}                                           & \multicolumn{4}{c|}{Weighted} & \multicolumn{4}{c|}{Mean}     & \multicolumn{4}{c|}{Median}   & \multicolumn{4}{c|}{Quantile} \bigstrut[t]\\
      &       & A     & B     & C     & D     & E     & F     & G     & H     & I     & J     & K     & L     & M     & N     & O     & P     & Q     & R     & S     & T     & U     & V     & W     & X     & Y \bigstrut[b]\\
\hline
\end{tabular}%

		\normalsize
		\caption{Overall payoffs to animal $A$ in parameter regime III when it pursues the strategy in the row and animal $B$ pursues the strategy in the column. The letters in the rows and columns correspond to the strategies in Table \ref{T:LearningStratsTable}.
			The parameter values are $c_W = 0.1$, $c_L = 1.2$, $\gamma = 0.995$, and $k = 1$.
			We colour cells as in Table \ref{T:RegimeILearningColourOnly}.}
		\label{T:RegimeIIILearningColourOnly}
	\end{table}
\end{landscape}



\end{document}